\PassOptionsToPackage{unicode}{hyperref}
\PassOptionsToPackage{hyphens}{url}
\PassOptionsToPackage{dvipsnames,svgnames,x11names}{xcolor}
\documentclass[12pt]{article}
  
\usepackage{graphicx}  
\usepackage{amsthm}
\usepackage{amsmath,amssymb}
\usepackage{iftex}
\usepackage{graphicx,psfrag,epsf}
\usepackage{enumerate}
\usepackage{natbib}
\usepackage{url} 
\usepackage{bm}
\usepackage{algorithm}
\usepackage{xcolor}
\usepackage{mathtools}
\usepackage{soul}
\usepackage{comment}
\usepackage{multirow}

\usepackage[colorinlistoftodos,textwidth=3.5cm, textsize=tiny]{todonotes}

\newcommand{\tik}{ {\color{blue} ${\cal \checkmark}$} }
\newcommand{\x}{ {\color{red}$\boldsymbol{\times}$} }

\newcommand{\ESBFhat}{E\widehat{SB}F}

\newtheorem{thm}{Theorem}
\newtheorem{rem}{Remark}

\newtheorem{defn}{Definition}

\ifPDFTeX
  \usepackage[T1]{fontenc}
  \usepackage[utf8]{inputenc}
  \usepackage{textcomp} 
\else 
  \usepackage{unicode-math}
  \defaultfontfeatures{Scale=MatchLowercase}
  \defaultfontfeatures[\rmfamily]{Ligatures=TeX,Scale=1}
\fi
\usepackage{lmodern}
\ifPDFTeX\else  
\fi
\IfFileExists{upquote.sty}{\usepackage{upquote}}{}
\IfFileExists{microtype.sty}{
  \usepackage[]{microtype}
  \UseMicrotypeSet[protrusion]{basicmath} 
}{}
\makeatletter
\@ifundefined{KOMAClassName}{
  \IfFileExists{parskip.sty}{%
    \usepackage{parskip}
  }{
    \setlength{\parindent}{0pt}
    \setlength{\parskip}{6pt plus 2pt minus 1pt}}
}{
  \KOMAoptions{parskip=half}}
\makeatother
\usepackage{xcolor}
\makeatletter
\ifx\paragraph\undefined\else
  \let\oldparagraph\paragraph
  \renewcommand{\paragraph}{
    \@ifstar
      \xxxParagraphStar
      \xxxParagraphNoStar
  }
  \newcommand{\xxxParagraphStar}[1]{\oldparagraph*{#1}\mbox{}}
  \newcommand{\xxxParagraphNoStar}[1]{\oldparagraph{#1}\mbox{}}
\fi
\ifx\subparagraph\undefined\else
  \let\oldsubparagraph\subparagraph
  \renewcommand{\subparagraph}{
    \@ifstar
      \xxxSubParagraphStar
      \xxxSubParagraphNoStar
  }
  \newcommand{\xxxSubParagraphStar}[1]{\oldsubparagraph*{#1}\mbox{}}
  \newcommand{\xxxSubParagraphNoStar}[1]{\oldsubparagraph{#1}\mbox{}}
\fi
\makeatother

\usepackage{longtable,booktabs,array}
\usepackage{calc} 
\usepackage{etoolbox}
\makeatletter
\patchcmd\longtable{\par}{\if@noskipsec\mbox{}\fi\par}{}{}
\makeatother
\IfFileExists{footnotehyper.sty}{\usepackage{footnotehyper}}{\usepackage{footnote}}
\makesavenoteenv{longtable}
\usepackage{graphicx}
\makeatletter
\def\maxwidth{\ifdim\Gin@nat@width>\linewidth\linewidth\else\Gin@nat@width\fi}
\def\maxheight{\ifdim\Gin@nat@height>\textheight\textheight\else\Gin@nat@height\fi}
\makeatother
\setkeys{Gin}{width=\maxwidth,height=\maxheight,keepaspectratio}
\makeatletter
\def\fps@figure{htbp}
\makeatother

\makeatletter
\@ifpackageloaded{caption}{}{\usepackage{caption}}
\AtBeginDocument{%
\ifdefined\contentsname
  \renewcommand*\contentsname{Table of contents}
\else
  \newcommand\contentsname{Table of contents}
\fi
\ifdefined\listfigurename
  \renewcommand*\listfigurename{List of Figures}
\else
  \newcommand\listfigurename{List of Figures}
\fi
\ifdefined\listtablename
  \renewcommand*\listtablename{List of Tables}
\else
  \newcommand\listtablename{List of Tables}
\fi
\ifdefined\figurename
  \renewcommand*\figurename{Figure}
\else
  \newcommand\figurename{Figure}
\fi
\ifdefined\tablename
  \renewcommand*\tablename{Table}
\else
  \newcommand\tablename{Table}
\fi
}
\@ifpackageloaded{float}{}{\usepackage{float}}
\floatstyle{ruled}
\@ifundefined{c@chapter}{\newfloat{codelisting}{h}{lop}}{\newfloat{codelisting}{h}{lop}[chapter]}
\floatname{codelisting}{Listing}

\makeatother
\makeatletter
\@ifpackageloaded{caption}{}{\usepackage{caption}}
\@ifpackageloaded{subcaption}{}{\usepackage{subcaption}}
\makeatother

\ifLuaTeX
  \usepackage{selnolig}  
\fi
\usepackage[]{natbib}
\usepackage{bookmark}

\IfFileExists{xurl.sty}{\usepackage{xurl}}{} 
\hypersetup{
  pdftitle={Title},
  pdfauthor={Author 1; Author 2},
  pdfkeywords={3 to 6 keywords, that do not appear in the title},
  colorlinks=true,
  linkcolor={blue},
  filecolor={Maroon},
  citecolor={Blue},
  urlcolor={Blue},
  pdfcreator={LaTeX via pandoc}}

\newcommand{\anon}{1}

\begin{document}

\def\spacingset#1{\renewcommand{\baselinestretch}%
{#1}\small\normalsize} \spacingset{1}


\if1\anon
{
  \title{\bf Stochastic Bayes factors: why, when, and how}
\author{ \begin{minipage}[t]{0.45\textwidth} \centering \textbf{Leonardo Egidi}\\ Department of Economics, Business, Mathematics,\\ and Statistics ``Bruno de Finetti'',\\ Via Valerio 4/1D, University of Trieste, Italy\\ ORCID: 0000-0003-3211-905X\\ Mail: legidi@units.it \end{minipage} \hspace{0.05\textwidth} \begin{minipage}[t]{0.45\textwidth} \centering \textbf{Ioannis Ntzoufras}\\ Department of Statistics,\\ Athens University of Economics and Business,\\ 28is Oktovriou 76, Athina 104 34, Greece\\ ORCID: 0000-0002-7615-0334\\ Mail: ntzoufras@aueb.gr \end{minipage} \thanks{ Leonardo Egidi is associate professor of statistics, Department of Economics, Business, Mathematics, and Statistics \lq\lq Bruno de Finetti\rq\rq{}, University of Trieste, Trieste, Italy, mail: legidi@units.it. Ioannis Ntzoufras is full professor of statistics, Department of Statistics, Athens University of Economics and Business, Athens, Greece, mail: ntzoufras@aueb.gr} }

  \maketitle

} \fi

\if0\anon
{
  \bigskip
  \bigskip
  \bigskip
  \begin{center}
    {\LARGE\bf Stochastic Bayes factors: why, when, and how}
\end{center}
  \medskip
} \fi

\bigskip
\begin{abstract}
The Bayes factor (BF) is a central tool in Bayesian hypothesis testing and model selection, yet its practical use is often challenged. Classical BFs depend heavily on prior specification, cannot be applied with improper priors, and are typically interpreted through arbitrary evidence scales. Moreover, they fail to capture uncertainty inherent in the data, leading to an analogy with frequentist $p$-values, and primarily reflect prior-predictive rather than posterior-predictive performance.
We introduce the stochastic Bayes factor (SBF), a new framework that extends the BF by explicitly incorporating uncertainty via replicated data. Formally, the SBF is defined as a push-forward measure transferring the BF from the observed data space to that of replications. This approach generalizes previous calibration proposals, while emphasizing posterior-predictive replication as a robust alternative. We establish key theoretical properties, including model consistency, compatibility and dominance, ensuring that SBFs preserve desirable Bayesian guarantees.

An algorithmic routine is then proposed to operationalize the SBF, guiding model discrimination in a principled way while naturally providing model calibration. Simulation studies and real applications confirm that the SBF offers improved robustness and predictive reliability compared to the classical BF, by providing a valuable tool for model comparison.
\end{abstract}

\noindent%
\vfill

\spacingset{1.2} 

\section{Introduction}
\label{sec:intro}

In the modern Bayesian protocol the Bayes factor \citep[BF;][]{kass1995bayes} still represents a milestone for model selection and hypothesis testing.  Given some data $\bm{y}=(y_1, y_2, \dots, y_n)$ and two competing parametric statistical models $\mathcal{M}_1, \mathcal{M}_2$, the Bayes factor $BF_{12}(\bm{y})$ is defined as the ratio between the two models' marginal likelihoods, and  measures the  strength of the evidence favoring $\mathcal{M}_1$ against $\mathcal{M}_2$ for the given sample; equivalently, the BF is the multiplicative factor applied to the prior odds to determine the posterior odds of the two competing models. 
A $BF_{12}(\bm{y})$ value greater than one ($<1$) suggests that the data provide higher support from model $\mathcal{M}_1$ ($\mathcal{M}_2$) than model $\mathcal{M}_2$ ($\mathcal{M}_1$), and the posterior odds in favor of $\mathcal{M}_1$ will be greater (smaller) than the prior odds.

\cite{jeffreys1961theory}, \cite{kass1995bayes} and others have formulated a scale of evidence for the observed BF. In particular,  
the interpretation of $BF_{12}(\bm{y})$ is based on the threshold values $(1,3,10,30,100, +\infty)$, which define intervals where evidence in favor of  $\mathcal{M}_1$ is typically characterized as 
anecdotal, moderate, strong, very strong, and extreme, respectively. 
Evidence in favor of  $\mathcal{M}_2$ can be assessed symmetrically through the reciprocal Bayes factor,  $BF_{21}(\bm{y}) = 1/BF_{12}(\bm{y})$.

Although its outstanding relevance in Bayesian model selection and hypothesis testing for discriminating between two candidate models or two alternative hypotheses, its application is limited and constrained by some  `philosophical' and computational issues. We list here the main points of discussion surrounding the BF ecosystem.

\begin{enumerate}
\item \emph{Use of weak prior information, or improper priors}.
BFs cannot be used with improper prior distributions \citep{o1995fractional, berger1996intrinsic}: in fact, if $p(\bm{\theta}) = k \times h(\bm{\theta})$, where $h(\cdot)$ is a function whose integral over the parameter space diverges, and $k$ is an  unspecified normalizing constant, the BF depends on a subjective $k$.  Partial \citep{lempers1971posterior}, fractional \citep{o1995fractional}, and intrinsic \citep{berger1996intrinsic}  BF variants have been proposed to fix the issue.

\item \emph{Scale of evidence and lack of uncertainty}.  The scale of evidence proposed by \cite{jeffreys1961theory} is empirical and subjective, without any clear methodological justification 
\citep{held2018p}.

\item \emph{Interpretation: measure of prior-predictive performance}.
As brilliantly remarked by some scholars, such as \cite{carpenter2022bayes}, Bayes factors measure prior predictive performance rather than posterior-predictive performance. 
Nevertheless, alternatives such as the posterior Bayes factors \citep{aitkin1991posterior}, where, in brief, the posterior distribution is used in place of the prior, do not represent a valid alternative, since they use the data twice and therefore violate many principles supporting the Bayesian paradigm while not completely solving the issues above.

\item \emph{Prediction and connection with LOO}.  
Relevant to the previous point is the fact that the BF fails to capture the predictive accuracy of the models under comparison. However, there is a strong connection of leave-one-out (LOO) to what \cite{gelfand1994bayesian} called the \emph{pseudo-Bayes factor} (PSBF) which they attribute to \cite{geisser1979predictive}, where each model marginal likelihood is replaced by the product of the leave-one-out predictive densities. Another connection between the marginal likelihood---and by consequence, BFs---and cross-validation in a predictive fashion is revealed by \cite{fong2020marginal}. A purely predictive version of BF is instead proposed by \cite{trotta2007forecasting}.

\item \emph{Prior-predictive calibration}. \cite{garcia2005calibrating} proposed to calibrate the BF with the prior-predictive distribution in order to obtain more robust and reliable results: in brevity, they suggested to use the calibrated distributions of the BF from both the models under investigation, and then select that model for which the observed BF is less surprising. This proposal allows to include in the framework the randomness in the data and, then, to account for uncertainty.

\end{enumerate}

From the aforementioned points, it emerges how the BF represents a valid but somehow controversial tool to perform model selection in the Bayesian framework. 
To this line, in this paper, we formally define and introduce the notion of \emph{stochastic Bayes factor} (SBF) to explicitly unify and formalize the approaches indicating that it is necessary to quantify and account for the BF uncertainty.  
The definition of SBF relies on the generation of data replications from an appropriate model distribution and then the evaluation of the BF on these replicants: in formal measure's theory words, the SBF is then a \emph{push-forward measure} \citep{bogachev2007measure} transferring the BF from the space of observed to that of replicated data. 

The idea of sampling hypothetical data and consequently evaluating the BF is not merely new.
\cite{garcia2005calibrating} sampled from the prior-predictive distribution, whereas  \cite{schad2022workflow} proposed a \emph{simulation-based calibration} approach  to compute Bayes factors on replicated posterior-predictive datasets and assess how variable are the results across different simulations of the same study. 
\cite{robert2014} proposed to look and concentrate on the posterior predictive distribution of the Bayes factors rather than the prior-predictive one. 
In line with the latter authors, we argue that the use of posterior-predictive samples provides a robust and effective alternative for SBFs. 
To motivate this choice, we establish theoretical properties such as model consistency, compatibility, and dominance. 
Moreover, through practical examples, we show that posterior-predictive calibration more effectively recovers the true data-generating mechanism in an $\mathcal{M}$-closed setting \citep{yao2018using}, where the true model is assumed to belong to the model space $\mathcal{M} = \{\mathcal{M}_1, \mathcal{M}_2, \ldots, \mathcal{M}_q\}$.

To fully provide a practical tool for model discrimination, we incorporate the aforementioned SBFs in an algorithmic routine to 
effectively drive selection choices between competing models.
We feel this proposal could be fruitful in many  settings and applications, with the desirable property of acknowledging a measure of uncertainty for the two models and automatically accomplish a sort of model calibration in terms of the ppd.
 
The remainder of the paper is organized as follows. Section~\ref{sec:sbf} introduces the general notion of stochastic BF. Section \ref{sec:novel} presents key theoretical properties of stochastic Bayes factors while Section \ref{sec:algo} provides an algorithmic procedure, based on the use of SBFs, for discriminating between two competing models. In Section~\ref{sec:ex}, the behavior of the SBF is examined through simulated examples and real-data applications. Section~\ref{sec:disc} concludes the paper.

\section{Definition and computation of stochastic Bayes factors}
\label{sec:sbf}

Once a Bayes factor  has been computed from the observed data $\bm{y}$, one needs to assess its strength in order to perform pairwise model comparisons. 
However, treating the BF as fixed after observing the data does not seem appropriate since it ignores uncertainty coming from the fact that it is estimated from a random sample, as remarked by \cite{garcia2005calibrating}. 
Hence, the Bayes factor $BF_{\ell k}(\bm{Y})$ itself, under repeating data sampling, can be considered a random variable following its own (sampling) distribution. 
From this (sampling) distribution of BF, it is  possible to derive theoretical properties which can be used to measure the agreement between the observed value of BF and each of the two models under investigation. 
Thus, it seems natural to calibrate the Bayes factor according to the data expected under each model under consideration and  take this into account  in our final model evaluation.
The concept of calibration in this context  is analogous to the classical hypotheses tests, where the decision rule is  based on the sampling distribution of the test statistic.

To clarify the role of the uncertainty in the BF derivation, we introduce the notion of \emph{stochastic Bayes factor} (SBF), denoted by  $SBF_{\ell k}(\bm{Y})$.  The SBF is defined as the BF evaluated on random data $\bm{Y}$ generated when the model distribution $p_{\tau}(\bm{y}), \tau=\ell \vee k$ is considered the true data generating  mechanism. 

Although the ``true'' data generating  mechanism is typically unknown in practical applications, we adopt it here as a conceptual device, following the approach of \cite{garcia2005calibrating}.

Assuming then that we can identify the model $\mathcal{M}_\tau$, with $\tau=\ell \vee k$, as the ``true'' data generating  mechanism, we define the \emph{expected SBF} as the average SBF 
with respect to the true model distribution:

\begin{equation}
	ESBF_{\ell k} 
 = \int BF_{\ell k}(\bm{y}^*)p_\tau(\bm{y}^*)  \nu( d \bm{y}^*) = \text{E}_{p_\tau(\bm{y}^*)}[SBF_{\ell k}(\bm{y}^*)],
\label{eq:esbf}
\end{equation}
where $\nu$ is a support measure defined on the sample space $\mathcal{Y}$. 
In the above notation and hereafter, we use the symbol $\bm{y}^*$  to denote the replicated data coming from $p_\tau(\bm{y}^*)$ to separate them from the observed data $\bm{y}$. 
The  ESBF in~\eqref{eq:esbf} is the expected value of the SBF with respect to the 
true model distribution;  thus, $p_\tau$ could be any reasonable distribution for the observables, such as the prior-predictive, the posterior-predictive, or the sampling distribution. 
At this stage, we are ready to provide a formal definition of the SBFs under prior/posterior-predictive replications.
\begin{defn}[Stochastic BF]
	\label{def:sbf}
	The SBF of model $\mathcal{M}_\ell$ versus model $\mathcal{M}_k$ may be formally defined as a {push-forward measure} \citep{bogachev2007measure}, denoted by 
	\begin{equation}
		{\mathcal Q}_{\ell k|\tau} \coloneqq BF_{\ell k}(\bm{Y})  \# PP_\tau (\bm{Y})  ~\mbox{for}~\tau=\ell \vee k, 
		\label{push_forward}
	\end{equation}
	with $PP_\tau$ denoting the prior or posterior predictive distribution of model $\mathcal{M}_\tau$: the SBF is then the distribution of $BF(\bm{Y})$ when $\bm{Y}$ is generated through the prior or posterior predictive distribution of the model. In the following, we will use the symbol $Q$ to denote the random variable $Q \sim \mathcal{Q}$. 
\end{defn}

If $p_\tau$ is the prior-predictive distribution, then the SBF coincides with the calibrated proposal of \cite[Section 4]{garcia2005calibrating}. As far as we know, this is the first attempt to formally define expected BFs computed from sampled replications, even if they are fruitfully adopted in the SBC context by \cite{schad2022workflow}. 
Under this scenario, the ESBF can be approximated by using $T$ samples from  the prior-predictive distribution, which yields a sort of simple/\emph{naive} Bayes estimator \citep{hammersley2013monte}.
%

However, we may argue that the evaluation of a Bayes factor, and ultimately the comparison of two models, should be based on the posterior-predictive distribution (ppd) rather than the prior-predictive. Indeed,  \cite{robert2014} remarks that this choice allows to:
(a) concentrate in a more relevant region of the parameter space, which is crucial in weakly-informative frameworks \citep{gelman2008weakly}; 
(b) reproduce the behavior of the BF for replicated values  of the observations similar to the original observations $\bm{y}$;
(c) go beyond the indeterminacy of the BF arising from the use of improper priors; 
(d) avoid any issue about the \lq\lq double use of the data\rq\rq{}, since,  as long as the evaluation is not used to reach a decision, the aim is just to produce an estimator of the posterior loss.

This suggestion translates in choosing the posterior predictive density, 
$m(\bm{y}^*|\bm{y})$,
in place of the prior-predictive, $m(\bm{y}^*)$,
as the true model distribution in Equation \eqref{eq:esbf}. Under this alternative scenario, the ESBF can be approximated by the following expression:

\begin{equation}
\ESBFhat_{\ell k} = \frac{1}{T} \sum_{t=1}^{T} {BF_{\ell k}({\bm{y}}^{*(t)}_\tau)}, 
\ \ \underbrace{
	{\bm{y}}^{*(t)}_\tau \sim p_\tau({\bm{y}}^*|\tilde{\theta}^{(t)}_\tau),
\ \ \tilde{\theta}^{(t)}_\tau \sim p_\tau(\theta_\tau|\bm{y}), }_{\mbox{\scriptsize\it samples from the posterior-pred. distribution}} 
\ \tau =\ell \vee k,
\label{eq:est_esbf2}
\end{equation}
where each ${\bm{y}}^{*(t)}_\tau$ is a posterior-predictive sample from the true model $\mathcal{M}_\tau$, and  $T$ is the sample size for data replications.

To sum up, Equation \eqref{push_forward} allows us to introduce alternative estimators for expected SBFs. The  expected value is taken with respect to either the prior-predictive or the posterior-predictive density among other choices (see Eq. 
\ref{eq:est_esbf2} for the latter). 
In the next sections we provide some theoretical properties for the aforementioned quantities.

\section{Model discrimination properties of stochastic Bayes factors}
\label{sec:novel}
In this section, we use SBFs to introduce an innovative algorithmic approach for model discrimination (Section \ref{sec:algo}).
To this end, we first introduce a set of theoretical definitions and properties, together with some mild mathematical conditions.
The final deliverable will be a sequence of steps aimed to the effective discrimination between the competing models $\mathcal{M}_\ell$ and $\mathcal{M}_k$. 

Let us assume the typical case of pairwise model comparison between ${\cal M}_1$ and ${\cal M}_2$.
Under this framework, we obtain the observed Bayes factor $BF_{12}$ and characterize its distribution under the (posterior or prior) predictive distribution of each model, denoted by $\mathcal{Q}_{12\mid 1}$ and $\mathcal{Q}_{12\mid 2}$, respectively.
Then, in the remainder of this section, we need to introduce the following notions which may characterize the model comparison between the two models:

\begin{itemize}
	\item \emph{Consistency}: It is the property of a model selection procedure which converges to the selection of the true model with probability equal to one when the data are generated from it - see Section \ref{sec:cons}; 
	

	\item \emph{Compatibility}: It is the case when the distribution of the SBF under a model is in agreement with the observed BF - see Section \ref{sec:protocol};
	
	
	
	\item \emph{Dominance}: We have dominance when all obtained Bayes factors ($BF_{12}$, $Q_{12|1}$ and $Q_{12|2}$) clearly support one of the two models - see Section \ref{sec:dom}.
	

\end{itemize}

From an operational standpoint, we will always assume to generate hypothetical data from both $\mathcal{M}_\ell$ and $\mathcal{M}_k$, thus to deal with the two corresponding distributions of SBFs, $\mathcal{Q}_{12|1}$ and $\mathcal{Q}_{12|2}$, respectively.

\subsection{Consistency of SBFs}
\label{sec:cons}

Model selection consistency \citep{liang2008mixtures}, or simply consistency \citep{kass1995bayes, walker2004new}, represents the well-known property of a Bayesian statistical procedure to recover the true model (or hypothesis) as the sample size grows.
As Bayesian statisticians, we need to ask ourselves whether the proposed ESBFs through Equation \eqref{eq:esbf} are \emph{consistent}, meaning that 
$$
ESBF_{\ell k} \rightarrow \infty ~(0) \mbox{ as } n \rightarrow \infty \mbox{ if } \mathcal{M}_\ell (\mathcal{M}_k) \mbox{ is the true model}. 
$$
Model consistency through BFs is a key property in the Bayesian setting to discriminate between two competing models, and should be always checked from a theoretical standpoint since it may be considered as the minimal requirement for a reasonable model selection procedure. 
For this reason, it is worth to assess the limiting behavior of the ESBFs w.r.t. the distribution from which we generate replicated data. 

Closely aligned with BFs' consistency, the crucial notion of posterior consistency  has been proposed by \cite{schwartz1965bayes} and is usually assumed as a working technical condition. We refer to \cite{chib2016bayes} for a related detailed list of references. 

\begin{defn}[Weak posterior consistency] 
	Let us consider $\boldmath{\Theta}=\{ \boldmath{\theta}_\ell \in \boldmath{\Theta}_\ell, \mathcal{M}_\ell \in  \mathcal{M} \}$ which is the combined parameter and model space. 
	Assume that the true density distribution is denoted by $p_\tau(\cdot|\theta_\tau)$ with $\mathcal{M}_\tau\in \mathcal{M}$, where $\mathcal{M}$ is the model space. 
	Further assume that $P^{(n)}_\tau$ is the $n$-fold product measure corresponding to the true distribution.  
	The posterior distribution $p_\tau(\boldmath{\theta}|\bm{y})$, for $\boldmath{\theta}\in\boldmath{\Theta}$, is weakly-consistent  at the true density $p_\tau(\cdot|\theta_\tau)$ if for every neighborood $U$ of $p_\tau(\cdot|\theta_\tau)$, $p_\tau(U| \bm{y}) \rightarrow 1$  as $n \rightarrow \infty$ with $P^{(\infty)}_\tau$-probability 1. 
\label{def:weak_cons}
\end{defn}

\begin{rem}
Commonly, the neighborhoods of interest can be balls around $p_\tau$, defined as $U = \{p \in \mathcal{M}: d(p_\tau, p) \leq \epsilon  \}$. Common metrics $d$ include total variation or Hellinger. \cite{schwartz1965bayes} established the Kullback-Leibler property for the prior distribution as an important criterion for demonstrating weak consistency.
\end{rem}

Bayesian consistency is typically assessed with respect to the observed data. In \cite{schad2022workflow}, simulation-based calibration techniques are developed to evaluate the variability of Bayes factors using posterior-simulated datasets; however, no theoretical guarantee of consistency is provided in that setup. The following result establishes sufficient conditions for proving consistency of ESBFs.

\begin{thm}
Assume a sequence of iid data observations $\bm{y} = \{{y}_i\}_{i=1,\ldots,n}$  with true distribution law $y_i \sim p_\tau(\bm{y}|\theta_\tau^0)$; 
where $\theta_\tau^0 \in \Theta_\tau $ is the vector of true  values of model parameters $\theta_\tau$ of model $\mathcal{M}_\tau, \tau=\ell \vee k$. 
We denote with $\bm{y}^*=\{y^*_i  \}_{i=1,\ldots,n}$ data replications under the true model. \\ 
Assuming that the following conditions hold:
\begin{itemize}
\item[(i)] \underline{Weak posterior consistency} defined in Definition \ref{def:weak_cons} holds.



\item[(ii)] \underline{Laplace's approximation}: the following Laplace's approximation for the marginal likelihood of model $\mathcal{M}_\tau$ with $d_\tau$ parameters holds: 
$$
m_\tau(\bm{y}) \approx p_\tau(\bm{y}|\hat{\theta}_\tau) p_\tau(\hat{\theta}_\tau)(2\pi)^{d_\tau/2}|I(\hat{\theta}_\tau) |^{-1/2}n^{-d_\tau/2}
$$
for an appropriate estimate $\hat{\theta}_\tau$ of $\theta_\tau$; where  $I(\hat{\theta}_\tau)$ is the Fisher information matrix evaluated in $\hat{\theta}_\tau$,  
\end{itemize}
then: 
\begin{itemize}
\item[(I)] the prior-predictive $ESBF_{\ell k}$ is consistent if $\mathcal{M}_\ell$ is the true model, thus $\tau =\ell$, but is not consistent if $\mathcal{M}_k$ is true, thus $\tau =k$. 
\item[(II)] The posterior-predictive $ESBF_{\ell k}$ is consistent when $\tau=\ell \vee k$.
\end{itemize}
\label{eq:thm1}
\end{thm}

The proof can be found in the e-Appendix A.1 the supplementary material.

As an immediate consequence of condition $(i)$ in Theorem \ref{eq:thm1}, $m_{\tau}(\bm{y}^*|\bm{y} ) \rightarrow p_\tau(\bm{y}^*| \theta_\tau)$ for replicated data $\bm{y}^*$.
The Laplace's approximation in condition  $(ii)$ is obtained via a second-order Taylor expansion of the log-posterior density around a suitable interior point $\hat{\theta}_\tau$ (typically the maximum likelihood or maximum a posteriori estimator), leading to a Gaussian approximation whose normalizing constant can be computed in closed form. The validity of Laplace’s method relies on standard regularity conditions: the parameter space has fixed dimension $d_\tau$; the log-likelihood is twice continuously differentiable in a neighborhood of $\theta_\tau$; the Fisher information matrix $I(\theta_\tau)$ is positive definite at $\hat{\theta}_\tau$; the prior density $p_\tau(\theta_\tau)$ is continuous and strictly positive in a neighborhood of the true parameter value; and the estimator $\hat{\theta}_\tau$ is consistent and asymptotically normal with rate $n^{1/2}$. Under these assumptions, the approximation holds with relative error of order $O(n^{-1})$ as $n \to \infty$; see, for example,  \cite{tierney1986accurate}, \cite{kass1995bayes}, \cite{van2000asymptotic}. 
Remark \ref{rem1}, which follows, helps to clarify the relevance and interpretation of Theorem \ref{eq:thm1}.

\begin{rem}[Interpretation and implications]
	\label{rem1}
The theorem highlights a fundamental asymmetry between prior-predictive and posterior-predictive evidence in model comparison. Under weak posterior consistency and standard Laplace-type regularity conditions, posterior-predictive scoring rules inherit the consistency of the posterior distribution itself, yielding consistent model selection regardless of which model is true. 

In contrast, prior-predictive quantities remain sensitive to the choice of prior through the marginal likelihood normalization term, which induces an implicit complexity penalty that does not vanish asymptotically when the competing model is correctly specified. As a consequence, the prior-predictive $ESBF_{\ell k}$ is consistent only when the true model coincides with the reference model, but may fail to concentrate on the true model otherwise. This result formalizes the intuition that posterior-predictive approaches effectively condition on the observed data and thus eliminate prior-induced distortions, while prior-predictive methods retain a persistent dependence on prior mass allocation even in large samples.
\end{rem}

\subsection{Model compatibility using observed BF}
\label{sec:protocol}

Once a pair of competing models $\{\mathcal{M}_\ell,\mathcal{M}_k\}$ is defined, our focus is to generate $n$ data replications from them and then obtain the distribution of the BFs across the simulated samples, the SBFs defined in Section \ref{sec:sbf}. According to the arguments expressed in Section \ref{sec:cons}, at a first stage one should check whether the posterior-predictive draws yield to model consistency; then, in a second stage, one should assess whether they are \emph{surprising} in comparison with the BF computed on the original data, $BF_{\ell k}(\bm{y})$: for such a reason, this section provides a general notion of compatibility for the two competing models.

As remarked by \cite{schad2022workflow}, the SBFs could be quite variable across posterior-predictive simulations, and this could generate a sort of \lq\lq dance of Bayes factors\rq\rq{}, analogously as the \lq\lq dance of $p$-values\rq\rq{} \citep{cumming2014new}: in these cases, a dataset is not necessarily highly informative for drawing clear conclusions about the scientific hypotheses in question.
To prepare us in performing this assessment, we define the following quantities---here for model $\mathcal{M}_\ell$, but the same applies for $\mathcal{M}_k$:
\begin{align}
	\begin{split}
\text{pr}p^L_\ell = &\ \text{Pr}\Big( ~SBF_{\ell k}(\bm{y}^*) \leq BF_{\ell k}(\bm{y})~ \Big| \mathcal{M}_\ell \Big) = \ \text{Pr}(Q_{\ell k|\ell}\leq BF_{\ell k}(\bm{y}))   
\end{split}
\label{eq:p_value}
\end{align}
where $\text{Pr}(\cdot |\mathcal{M}_\ell)$ stands for the predictive distribution of $SBF_{\ell k}(\bm{y}^*)$ under the model $\mathcal{M}_\ell$, and the quantities $\text{pr}p^L_{j}$, $\text{pr}p^R_{j}$ denote the predictive $p$-values computed under the model $\mathcal{M}_\ell$ for the left and right tail, respectively---the same definitions apply under model $\mathcal{M}_k$. Then, as suggested by \cite{garcia2005calibrating}, we define the two-sided predictive $p$-values for the BFs under model $\mathcal{M}_\ell$ and $\mathcal{M}_k$, respectively:
\begin{align}
\text{pr}p_\ell = \text{min} \Big\{ \text{pr}p^L_\ell, 1-\text{pr}p^L_\ell\Big\},\ \  
\text{pr}p_k    = \text{min} \Big\{ \text{pr}p^L_k,    1-\text{pr}p^L_k \Big\}.
\label{eq:p_value2}
\end{align}
A candidate rule to assess compatibility for model $\mathcal{M}_\ell$---and analogously $\mathcal{M}_k$---is to check that the observed BF is \lq\lq plausible\rq\rq{} in terms of the SBF predictive distribution, i.e. it falls in the distribution of the SBF under model $\mathcal{M}_\ell$ ($\mathcal{M}_k$); this can be seen as a relaxation of  the global discrimination Rule 2 of \cite{garcia2005calibrating},  who proposed to select model $\mathcal{M}_\ell$ over $\mathcal{M}_k$ iff pr$p_\ell$ $>$ pr$p_k$. 
We note that the above condition is weaker, as it should be interpreted in terms of compatibility with a single model rather than model discrimination. We can therefore state the general definition of model compatibility.

\begin{defn}[Model $\epsilon$-compatibility]
\label{def:comp}
A model $\mathcal{M}_\tau$ is $\epsilon$-compatible in terms of SBF if its corresponding predictive $p$-value $\text{pr}p_\tau > \epsilon$, for a given $\epsilon >0$.
\end{defn}
%
%
The following theoretical result holds.

\begin{thm}[Asymptotic calibration of compatibility]
\label{thm:compatibility}
Assume that $\mathcal M_\tau$ is the true data-generating model, that weak posterior consistency defined in Definition \ref{def:weak_cons} holds for $p_\tau(\theta_\tau\mid\bm y)$, and that the predictive distribution of $SBF_{\ell k}(\bm y^*)$ under $\mathcal M_\tau$, with $\tau = \ell \vee k$, has a continuous cumulative distribution function. Then, for any $\epsilon\in(0,1/2)$,

$$
\Pr\left( {\cal M}_\tau \mbox{~is~} \epsilon\mbox{-compatible} \right)
\longrightarrow
1-2\epsilon,~~ \text{as } n\to\infty .$$
\end{thm}

The proof is contained in e-Appendix A.2 of the supplementary material.
Theorem \ref{thm:compatibility} implies that, under repeated sampling and for sufficiently large sample sizes, the true model will be $\epsilon$-compatible with high probability for small $\epsilon$-level. For instance, when $\epsilon$ = 0.01, the true model is $\epsilon$-compatible with probability $0.98$ for sufficiently large sample sizes.

\begin{rem}[Continuous CDF]
	\label{rem3}
The continuity assumption on $F_Q$ is satisfied whenever $\bm{y}^*$ follows 
a continuous sampling model. For discrete 
data-generating models, $F_Q$ is a step
function and the standard probability integral transform does not yield an exact $\text{Uniform}(0,1)$ limit. In such cases, an analogous result can be established using the randomized or mid-rank PIT \citep{czado2009predictive}, at the cost of introducing an auxiliary randomization. We leave this extension to future work.
\end{rem}

\subsection{Stochastic dominance}
\label{sec:dom}

Once consistency and compatibility between the two SBFs have been checked for two models under comparison, 
model discrimination can be further characterized by the concept of \emph{dominance}.
First, within the proposed framework, we define \emph{observed dominance} by examining the observed Bayes factor, interpreted according to a standard intuitive scale such as those proposed by \cite{jeffreys1961theory} and \cite{kass1995bayes}, and discussed in Section~\ref{sec:intro}.
Hence, the characterization of the Bayes factor can be now summarized as follows. 
When $BF_{12}(\bm{y}) \approx 1$, no dominance is observed (neither model dominates the other);
values between 1 and 3 indicate minor dominance of $\mathcal{M}_1$ over $\mathcal{M}_2$, between 3 and 10 indicate moderate dominance, between 10 and 30 strong dominance, between 30 and 100 very strong dominance, and values exceeding 100 indicate decisive (or full) dominance of $\mathcal{M}_1$ over $\mathcal{M}_2$.

Nevertheless, the innovation of this section is not simply the re-interpretation of the Bayes factor and the Bayesian evidence it provides
in favor of one model over another, 
but rather the introduction of \emph{stochastic dominance}. The notion of first-order  and second-order  \citep{hadar1969rules} stochastic dominance as a stochastic ordering between random variables in terms of their cumulative distribution functions (CDFs) is well-known in the probabilistic and decision theory literature. However, as far as we know from reviewing the statistical literature,  this concept is new when is defined in terms of the distributions of the Stochastic Bayes Factors evaluated under the (predictive) data-generating distributions of the two competing models.

Within this context, one model will be dominated over another one if the distribution of SBF under a data-generated distribution is clearly away from zero. 
Strictly speaking we need to specify the level of evidence $\varepsilon > 0$, but small enough, (ideally $\varepsilon \rightarrow  0^+$) in order to characterize the level of dominance. 

\begin{defn}[$\epsilon$-dominance] 
\label{def:dom1}
A model $\mathcal{M}_{\ell}$ is $\varepsilon$-dominant over $\mathcal{M}_{k}$ under model $M_\tau$, for $\tau \in \{\ell,k\}$ if 
\begin{equation}
p_{\ell k|\tau}^{d}= Pr(Q_{\ell k|\tau} < 1) \le \varepsilon 
\label{p_dominance}
\end{equation}
for some small $\varepsilon >0$; where $Q_{\ell k|\tau} \sim \mathcal{Q}_{\ell k|\tau}=BF_{{\ell k}}({\bf Y})\#PP_\tau({\bf Y})$ is the Stochastic BF of $\mathcal{M}_{\ell}$ against $\mathcal{M}_{k}$ evaluated with data generated under the data-generative (predictive) distribution of $\mathcal{M}_\tau$. 
\end{defn}

Of course, reversing the sequence of models in \eqref{p_dominance} will give us that if $p_{k \ell|\tau}^{d}= Pr(Q_{k \ell|\tau} < 1) =Pr(Q_{\ell k|\tau} > 1)\le \varepsilon$  then we can infer that model $\mathcal{M}_{k}$ is $\varepsilon$-dominant over $\mathcal{M}_{\ell}$ under model $\mathcal{M}_\tau$. 
In general, we can claim stochastic dominance in favor of one of the two models under comparison if 
\begin{equation}
D_{\ell k|\tau}= 2 \min \left\{ p_{\ell k|\tau}^{d},  \, p_{k \ell|\tau}^{d} \right\} \le 2 \varepsilon.
\label{eq:stoc_dom}
\end{equation}

From the above, $D_{\ell k|\tau}$ can take values from zero to one.
Alternatively, $(1-2\varepsilon)\times 100\%$ posterior intervals based on the $\varepsilon$ and $1-\varepsilon$ posterior quantiles can be used to visualize dominance. 
Following this approach, if zero is excluded from the $(1-2\varepsilon)\times 100\%$ interval, then we can infer $\varepsilon$-dominance in favor of one of the two competing models.
In practice, the value of $\tfrac{1}{2}D_{\ell k|\tau}$ can serve as the lowest $\varepsilon$-level of dominance of one of the two models for the SBF generated from $\mathcal{M}_\tau$. Distinct cases of full, clear, and mixed stochastic dominance are discussed in e-Appendix A.3 of the supplementary material.

Two additional cases can be observed where there is {\it no dominance} or the dominance is {\it inconclusive}.
No dominance will be observed when $D_{\ell k|\tau}$ is not small enough evidence, under both generating models ${\cal M}_\tau$ with $\tau \in \{\ell,k\}$, to support dominance of either of the two models under comparison. 
On the other hand, inconclusive dominance will occur when both SBFs support the opposite model from which the data have been generated, that is 
$p_{\ell k|\ell}^d<\epsilon$ and $p_{k \ell|k}^d<\epsilon$. Nevertheless, intuitively, we do not expect this kind of dominance to occur in practice unless something is really wrong with the specification of the two models under comparison. 

In conclusion, to determine which model is preferable, and to assess possible inconsistencies or misspecifications, we wish to consider:  \emph{strong evidence} (combination of dominance and compatibility), and  \emph{weak evidence} (dominance in the direction of support of the observed BF, but lack of compatibility.)
As a practical guideline, we will visually check compatibility and dominance through the 90\% posterior predictive credible intervals, by relegating pr$p$ and $p^d_{\ell k|\tau}$ values in the supplementary material.

\section{An SBF based algorithmic procedure for model discrimination}
\label{sec:algo}

Once the theoretical properties of consistency, compatibility, and dominance are investigated, we have a complete protocol for performing model selection via SBFs, as reported in Algorithm \ref{algo_bf}. In brief, under our procedure the user needs to: 
\begin{enumerate}
	\item generate SBFs from the predictive distributions of both competing models; 
	\item compute posterior summaries such as  ESBF and credible intervals; 
	\item assess the properties aforementioned through Sections \ref{sec:cons}--\ref{sec:dom}; and
	\item perform the final model evaluation and selection, if appropriate. 
\end{enumerate}

\begin{tiny}
\begin{algorithm}[H]
	\caption{Model comparison via stochastic Bayes factors}\label{algo_bf}
	\spacingset{1}
	\footnotesize
		\textbf{Inputs:}
		\begin{itemize}
			\item Observed data $\bm{y} = (y_1, \ldots, y_n) \overset{\text{iid}}{\sim} p(\bm{y} \mid \theta)$.
			\item Set of competing models $\mathcal{M} = \{\mathcal{M}_1, \mathcal{M}_2, \ldots, \mathcal{M}_q\}$.
			\item Pairwise observed Bayes factors $BF_{\ell k}(\bm{y})$ for each model pair $(\mathcal{M}_\ell, \mathcal{M}_k)$.
		\end{itemize}
		
		\textbf{Outputs:}  
		Posterior predictive distributions of stochastic Bayes factors (SBFs), estimated expected SBFs, and model selection outcomes.
		
		\vspace{0.2cm}
		\textbf{Algorithm:}
		
		\begin{enumerate}
			\item For each model $\mathcal{M}_\tau$, with $\tau \in \{\ell, k\}$:
			\begin{enumerate}
				\item For $t = 1, \ldots, T$:
				\begin{enumerate}
					\item Generate $\theta_\tau^{(t)} \sim p_\tau(\theta_\tau \mid \bm{y})$.
					\item Generate $\bm{y}_\tau^{*(t)} \sim p_\tau(\bm{y}^* \mid \theta_\tau^{(t)})$.
					\item Compute $BF_{\ell k}(\bm{y}^{*(t)}_\tau)$.
				\end{enumerate}
				\item Construct the empirical distributions of SBFs.
				\item Compute the estimated expected SBF (ESBF) as in Equation~\eqref{eq:est_esbf2}.
				\item Compute the posterior intervals of SBFs based on posterior quantiles.
				\item Compute the predictive $p$-values as in Equations~\eqref{eq:p_value}--\eqref{eq:p_value2}.
			\end{enumerate}
			\item Assess:
			\begin{itemize}
				\item Model compatibility (Definition~\ref{def:comp}) using the $p$-values in 1(d) and/or posterior intervals.
				\item Model dominance (Definition~\ref{def:dom1} and Equation \ref{eq:stoc_dom}).
			\end{itemize}
			\item Perform the final model selection: start checking compatibility, and then dominance. 
			\begin{itemize}
				\item  If a model is compatible and dominant $\Rightarrow$ the dominant model is selected (strong evidence).
				\item If no model is compatible, but one  is dominant $\Rightarrow$ the dominant model is selected (weak evidence).   
				\item If both models are compatible, but no model is dominant, then no clear evidence in favor of either model $\Rightarrow$ inconclusive model selection.
		\end{itemize}
		\end{enumerate}
\end{algorithm}
\end{tiny}

\begin{figure}
	\centering
	\includegraphics[width=6in,height=\textheight]{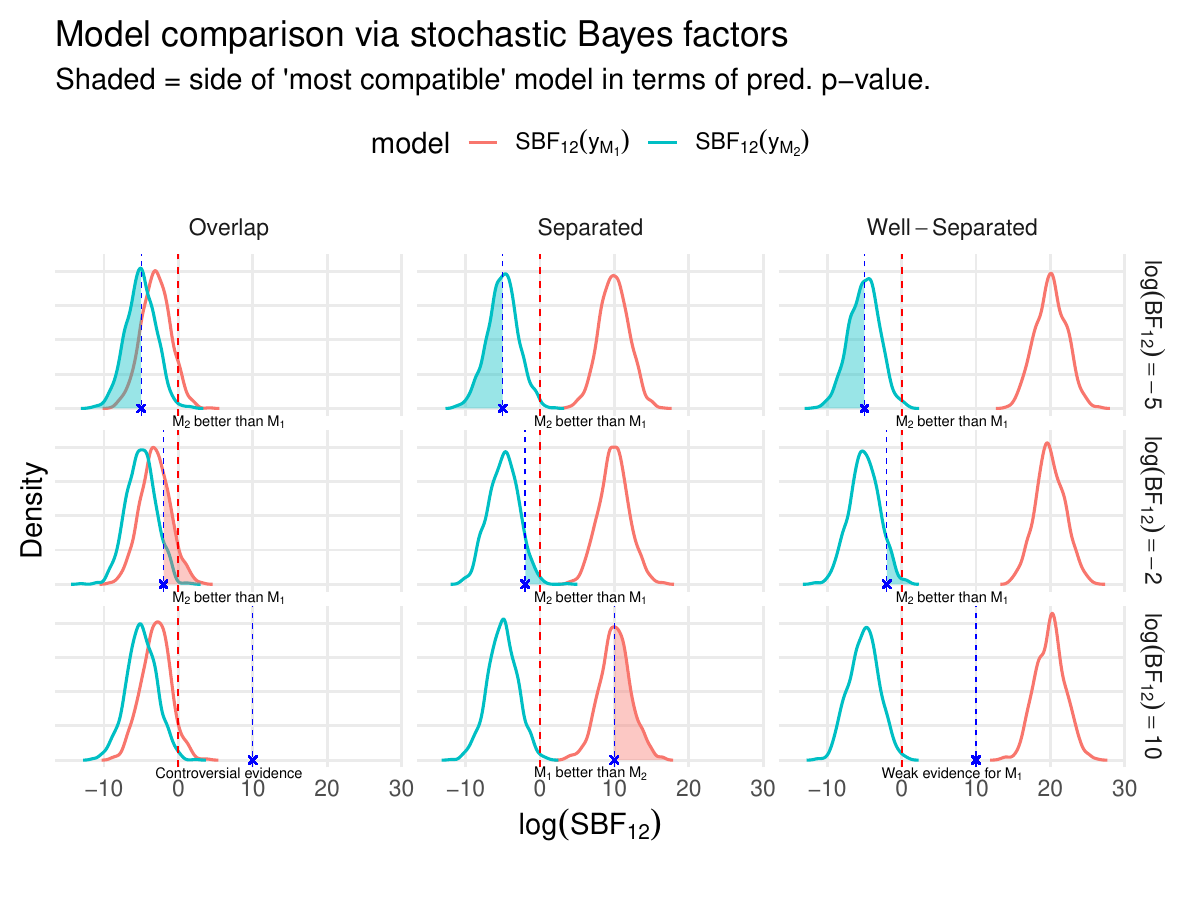}
	\parbox{0.7\textwidth}{\scriptsize \it Each column denotes a different scenario: (1) Overlap, (2) Separated and (3) Well Separated; 
		Each row denotes different values of the observed log--$BF_{12}(\bm{y})$;  
		Shaded colored areas: ``most compatible'' model (according to the prior-predictive $p$-values; see Eq. \ref{eq:p_value}--\ref{eq:p_value2}); 
		No colored areas: no compatibility for either model; 
		Dashed blue lines: observed log--$BF_{12}(\bm{y})$; 
		Dashed red line denotes zero (no model discrimination). Each panel reports a label indicating the best model.}
	\caption{Model discrimination between two competing models, $\mathcal{M}_1$ and $\mathcal{M}_2$, under three distinct scenarios.}
	\label{fig:panel}
\end{figure}

 Figure~\ref{fig:panel} displays logarithmic transformed SBFs obtained through posterior-predictive replications from two competing models, $\mathcal{M}_1$ and $\mathcal{M}_2$, $SBF_{12}(\bm{y}^*_{\mathcal{M}_1})$ and $SBF_{12}(\bm{y}^*_{\mathcal{M}_2})$ respectively, under three distinct artificial scenarios: (1) Overlap, (2) Separated and (3) Well Separated  (reported in columns) and three different values of the observed BF, $BF_{12}(\bm{y})$ (reported in rows). 
 The aim of this illustration is simply to summarize all possible (realistic or unrealistic) cases. 
 
 In brief, log-BF data under $\mathcal{M}_2$ are simulated from a $\mathcal{N}(-5,2^2)$ for all the three scenarios, whereas data under $\mathcal{M}_1$ are generated from $\mathcal{N}(-3,2^2)$, $\mathcal{N}(10,2^2)$, and $\mathcal{N}(20,2^2)$, respectively.   Shaded colored areas denote the \lq most compatible\rq{} model according to predictive $p$-values defined in Equations \eqref{eq:p_value}--\eqref{eq:p_value2}---see also the Rule 2 of \cite{garcia2005calibrating}---whereas no colored areas arise when there is no  compatibility for either model.
 The dashed blue line in each panel denotes the observed log-BF, $\log BF_{12}(\bm{y})$.

For simplicity, scrolling from left to right and from top to down, let us denote the nine panels in Figure \ref{fig:panel} through the symbols $C_{ij}$ for $i,j \in \{1,2,3\}^2$. 
The aim is to establish, for each single case, whether a clearly better model can be assumed and, consequently, whether model discrimination is meaningful. In the following, we evaluate dominance with $\varepsilon = 0.2$. Table \ref{tab:panel-cases} summarizes the results for all nine panels considered.

From Table \ref{tab:panel-cases}, we can conclude that 
$\mathcal{M}_2$ is better than $\mathcal{M}_1$ in all the six panels of the first two rows of Figure \ref{fig:panel}, from $C_{11}$ to $C_{23}$, and this is due to the fact that $\mathcal{M}_2$ is compatible across these six cases and $\varepsilon$-dominance is always guaranteed, either in a mixed or in a full sense---$\mathcal{M}_1$ is not dominant in the first column, being  $p^d_{12|1} = Pr(Q_{12|1} < 1) =  0.933 > \varepsilon = 0.2$, whereas $\mathcal{M}_2$ is dominant in the rest of the plot, being $p^d_{21|2} = Pr(Q_{21|2} > 1) =  0.006 < \varepsilon =0.2$.   Controversial evidence and lack of compatibility arise in panel $C_{31}$: the observed BF clearly favors $\mathcal{M}_1$, whereas both SBFs clearly support $\mathcal{M}_2$. 
This might  represent a rather unrealistic scenario in practice, which is not clear when it can arise, but we have included it in this ``toy'' illustration for completeness.  
Weak evidence for $\mathcal{M}_1$ appears in panel $C_{33}$, where neither model is compatible: $\mathcal{M}_1$, although clearly favored over $\mathcal{M}_2$, receives from the posterior predictive samples stronger support than would be expected, suggesting that the \lq true\rq{} model may lie somewhere between $\mathcal{M}_1$ and $\mathcal{M}_2$, despite much closer to $\mathcal{M}_1$ than $\mathcal{M}_2$. Finally, $\mathcal{M}_1$ is clearly favored in panel $C_{32}$, being compatible with respect to the log-BF equal to 10, an extremely high value of evidence, and $\epsilon$-dominant, albeit in a mixed sense.

\begin{table}
	\caption{Characterization of different model discrimination cases appearing in Figure \ref{fig:panel}.   \label{tab:panel-cases}}
	\begin{center}
	\scriptsize   
	\begin{tabular}{|c@{~~}c@{~~}c@{~~}c@{~~}|c@{~~}c@{~~}|c@{~~}c@{~~}|c|l|}
		\hline 
		&  & Obs. &  & \multicolumn{2}{c|}{${\cal M}_1$}& \multicolumn{2}{c|}{${\cal M}_2$}& Misspec.  &\\ 
		Case & Panel & log-BF & SBF  & C & D& C & D&   Diagn. & Conclusion \\
		\hline 
		\multirow{ 3}{*}{Overlap} 
		&$C_{11}$  &      $-5$         & $SBF_{12}$ & \tik & \x & \tik & \tik     &&  $\mathcal{M}_2$ is better than $\mathcal{M}_1$. \\
		&$C_{21}$  &$-2$& $SBF_{12}$   & \tik & \x & \tik & \tik   &&  $\mathcal{M}_2$ is better than $\mathcal{M}_1$.\\
		&$C_{31}$  &   $10$            & $SBF_{12}$ & \x & \x   & \x & \tik   &\x&  Controversial evidence.\\
		\cline{3-10}
		\hline
		\multirow{3}{*}{Separated} 
		&$C_{12}$  &     $-5$           &$SBF_{12}$ & \x & \tik & \tik   & \tik    && $\mathcal{M}_2$ is better than $\mathcal{M}_1$. \\
		&$C_{22}$  &$-2$ &$SBF_{12}$ &  \x & \tik   & \tik & \tik  && $\mathcal{M}_2$ is better than $\mathcal{M}_1$.   \\
		&$C_{32}$  &  $10$              &$SBF_{12}$ & \tik & \tik &   \x   & \tik  && $\mathcal{M}_1$ is better than $\mathcal{M}_2$.\\
		\cline{3-10}
		\hline
			\multirow{3}{*}{Well separated} 
		&$C_{13}$  &  $-5$              &$SBF_{12}$ & \x & \tik & \tik   & \tik    && $\mathcal{M}_2$ is better than $\mathcal{M}_1$. \\
		&$C_{23}$  &$-2$ &$SBF_{12}$ &   \x & \tik   & \tik & \tik  && $\mathcal{M}_2$ is better than $\mathcal{M}_1$.  \\
		&$C_{33}$  &  $10$              &$SBF_{12}$ & \x & \tik &  \x   & \tik  &\x& Weak evidence for $\mathcal{M}_1$.\\
		\cline{3-10}
		\hline 
		\multicolumn{10}{p{13cm}}{\scriptsize \it \textbf{\textit{ Misspec. Diagn.}}: Misspecification Diagnosis; \textbf{\textit{C}}: Compatibility; \textbf{\textit{D}}: Dominance; threshold $\varepsilon =0.2$ for both properties.  }
	\end{tabular}
\end{center}
\end{table}

We emphasize once again the introduction of gray areas of indecision, where the available evidence is not sufficient to clearly favour either of the two competing models ---see panels $C_{31}$ and  $C_{33}$. 
For instance, if the predictive distributions of the two SBFs support the model under which the predictive data were generated and the observed BF falls within the two separable distributions of SBFs, as in $C_{33}$, then no decisions could be taken in terms of global model selection. 
Similarly, if the distributions of SBFs support the generating model, but now the observed BF lies to the left or right of both distributions, then one model will appear clearly better than the other even if no compatibility arises.  
Nevertheless,  in this case the indicated best model is likely to be misspecified, as the unexpectedly stronger evidence might be driven by a model glitch.
An even more paradoxical case arises when the SBF distributions under both competing models support one model, while the observed BF favors the other (as in $C_{31}$). In this situation, no decision can be made, since the observed evidence is inconsistent with what is expected by either model specification.
These cases  are possibly related with model misspecification where  it is likely that we are in a $\mathcal{M}$-complete or even $\mathcal{M}$-open scenario, and the \lq correct\rq{} model is not a member of the model list.

\section{Applications}
\label{sec:ex}

	

\subsection{Nested linear regressions}
\label{sec:gprior-nested}

Assume a usual linear regression setup with 
$\bm{y} | {\cal M}_k \sim \mathcal{N}_n(  \bm{X}_k\bm{\beta}_k + \bm{\epsilon}, \sigma^2 I_n)$, 
where 
$\bm{X}_k$  is the $n\times p_k$  design matrix of full rank $p_k$,
$\bm{\beta}_k$ is the $p_k\times 1$ regression coefficient vector, and $I_n$ the $n \times n$ identity matrix. 

Let us now consider the case where we wish to compare two nested models ${\cal M}_1$ vs  ${\cal M}_2$ with parameters 
$\bm{\beta}_1=\bm{\beta}_{(1)}$ 
and $\bm{\beta}_2^T=(\bm{\beta}_{(1)}^T, \bm{\beta}_{(2)}^T)$, respectively. 
Model ${\cal M}_1$ can be obtained by considering the specification of model ${\cal M}_2$:  
$\bm{y} \sim \mathcal{N}_n(  \bm{X}_2\bm{\beta}_2 + \bm{\epsilon}, \sigma^2 I_n)$ by setting $\bm{\beta}_2^T=(\bm{\beta}_{(1)}^T, \bf{0}^T)$, that is $\bm{\beta}_{(2)}={\bf 0}$. 
Then, we can further consider the $g$-prior  \citep{zellner1986assessing} specification 
for the model regression parameters
as modified by \cite{liang2008mixtures} for the full based approach. 
Then the prior is given by  $\bm{\beta}_{(2)} \sim \mathcal{N}(\bm{0},  g (\bm{X}_2^T\bm{X}_2)^{-1} \sigma^2)$, with $p(\bm{\beta}_{(1)}, \sigma^2) \propto 1/\sigma^2$,  where  $g$ is the scaling factor controlling variable/model selection, and $\bm{X}_{(2)}$ is the $n \times p_2$ submatrix of $\bm{X}_2$ that corresponds to coefficients $\bm{\beta}_{(2)}$. 
Under this approach, the resulting Bayes factor for comparing model ${\cal M}_1$ vs  ${\cal M}_2$ is given by 
\begin{equation}
	BF_{12}(\bm{y}) = (1+ g)^{-(n-p_2-1)/2} \left[ 1+ g  \frac{1-R^2_2}{1-R^2_1}  \right]^{(n-p_1-1)/2},
	\label{eq:bf_2}
\end{equation}
where $R^2_1, R^2_2$ denote the coefficient of determination under the model $\mathcal{M}_1$ and $\mathcal{M}_2$, respectively.

Then, we generate $T$ posterior-predictive replicated datasets from each of the models under comparison, each of them following a multivariate student-$t$ distribution with $n$ degrees of freedom,
\begin{eqnarray*}
	\bm{y}^* | {\cal M}_k &\sim&  \ t_{n}(\bm{\mu}_k^{\text{pred}}, \bm{\Sigma}_k^{\text{pred}}), \mbox{~with~}  \\
	\bm{\mu}_k^{\text{pred}} &=& \omega \bm{X}_k \bm{\widehat{\beta}}_k, \\
	\bm{\Sigma}_k^{\text{pred}} &=& \left(\bm{I}+ \omega \bm{X}_k(\bm{X}_k^T\bm{X}_k)^{-1}\bm{X}_k^T \right) s^2, 
\end{eqnarray*}
where $\omega =g/(g+1)$, 
$s^2 = \frac{1}{n-p_k} \left( \sum_{i=1}^n ({y}_i -\hat{y}_i)^2 + \frac{1}{g+1} \widehat{\bm{\beta}}_k^T \bm{X}_k^T\bm{X}_k \widehat{\bm{\beta}}_k\right)$ is the sample variance estimate, $p_k$ is the dimension of model ${\mathcal M}_k$ for $k=1,2,3$, 
and $\widehat{\bm{\beta}}_k = (\bm{X}_k^T \bm{X}_k)^{-1}\bm{X}_k^T \bm{y}$ is the maximum  likelihood estimate for $\bm{\beta}_k$.


To implement our SBF procedure detailed in Algorithm \ref{algo_bf}, we  consider three nested linear regression models $\mathcal{M}_1 \subset\mathcal{M}_2 \subset \mathcal{M}_3$,  with $p_1, p_2$ and $p_3$ parameters, $p_1 < p_2< p_3$, respectively.

\begin{figure}
	\centering
	\includegraphics[width=2in,height=\textheight]{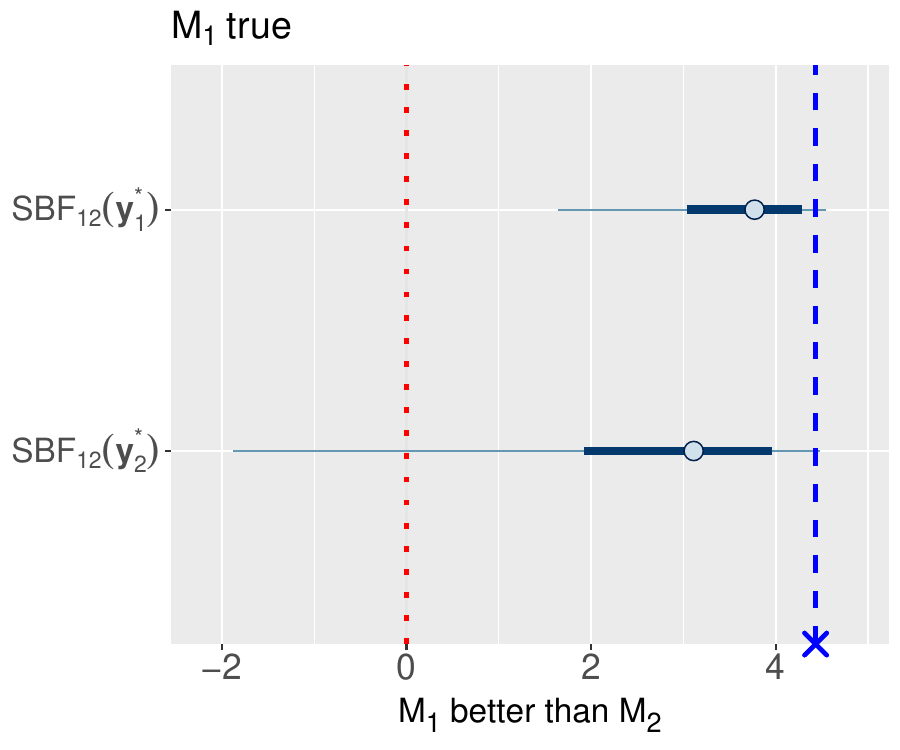}~
	\includegraphics[width=2in,height=\textheight]{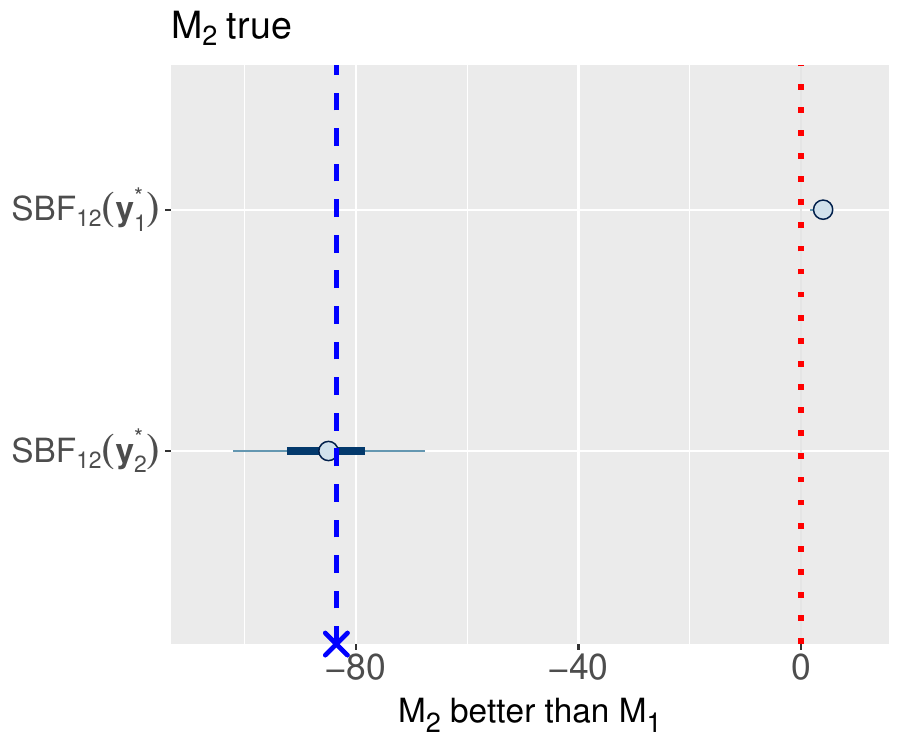}~
	\includegraphics[width=2in,height=\textheight]{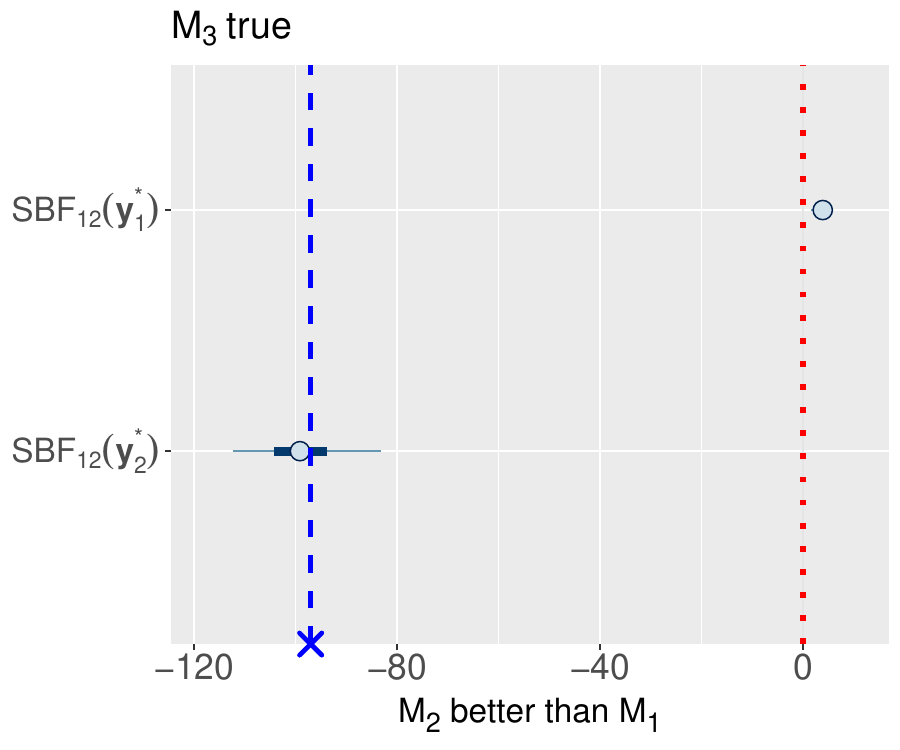}\\
	\includegraphics[width=2in,height=\textheight]{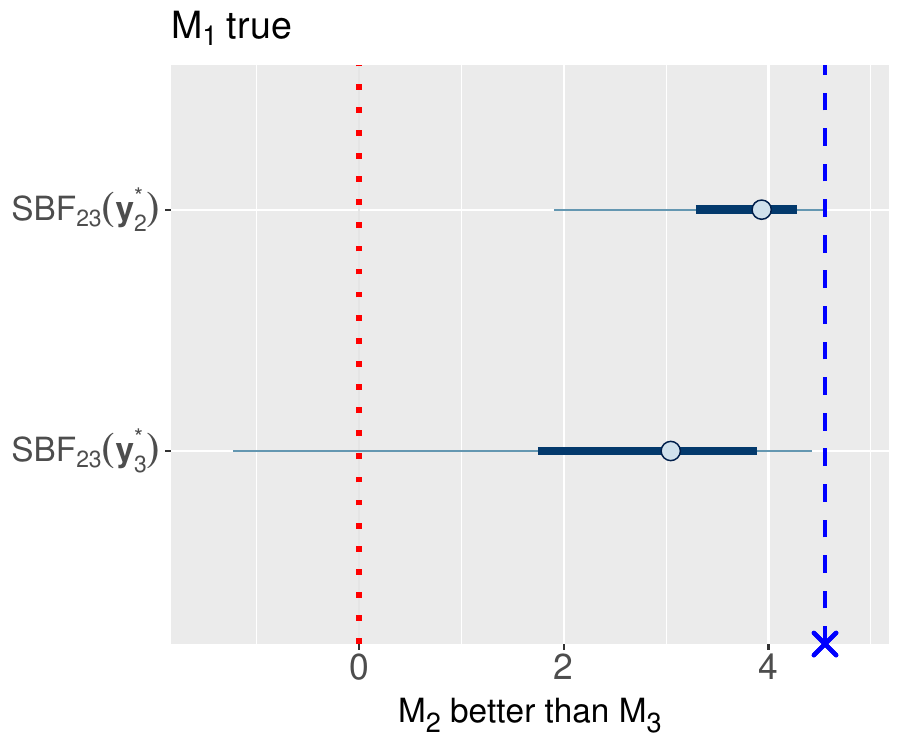}~
	\includegraphics[width=2in,height=\textheight]{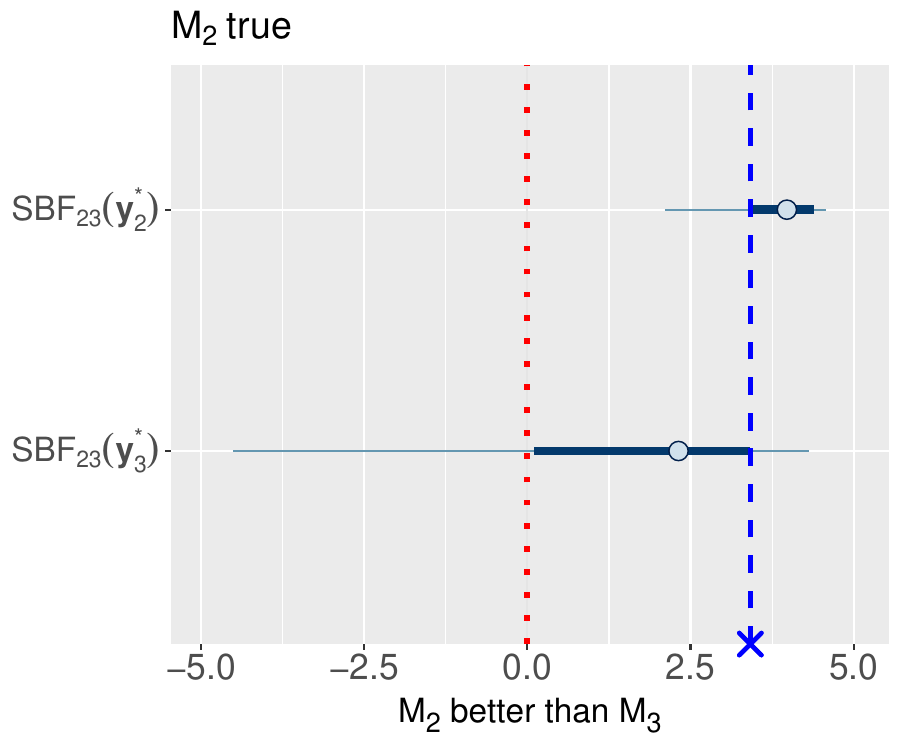}~
	\includegraphics[width=2in,height=\textheight]{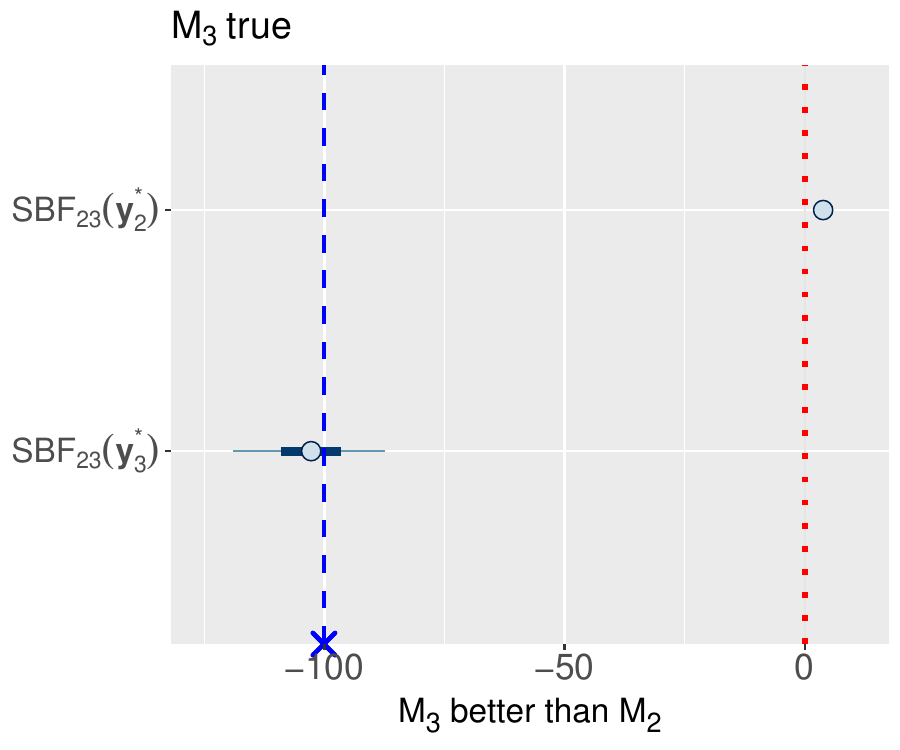}\\
	\includegraphics[width=2in,height=\textheight]{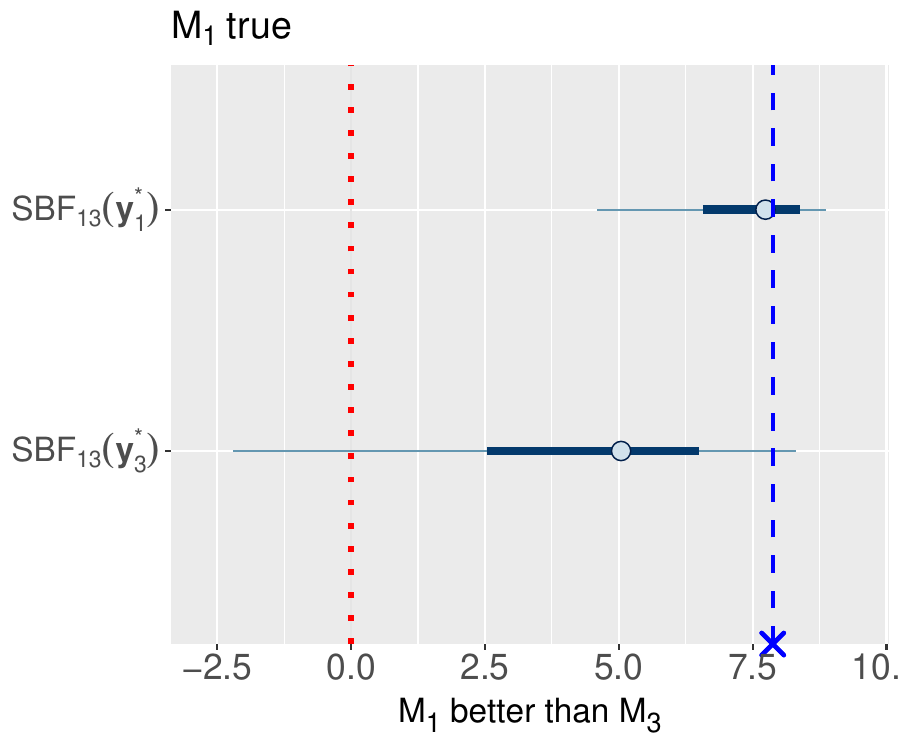}~
	\includegraphics[width=2in,height=\textheight]{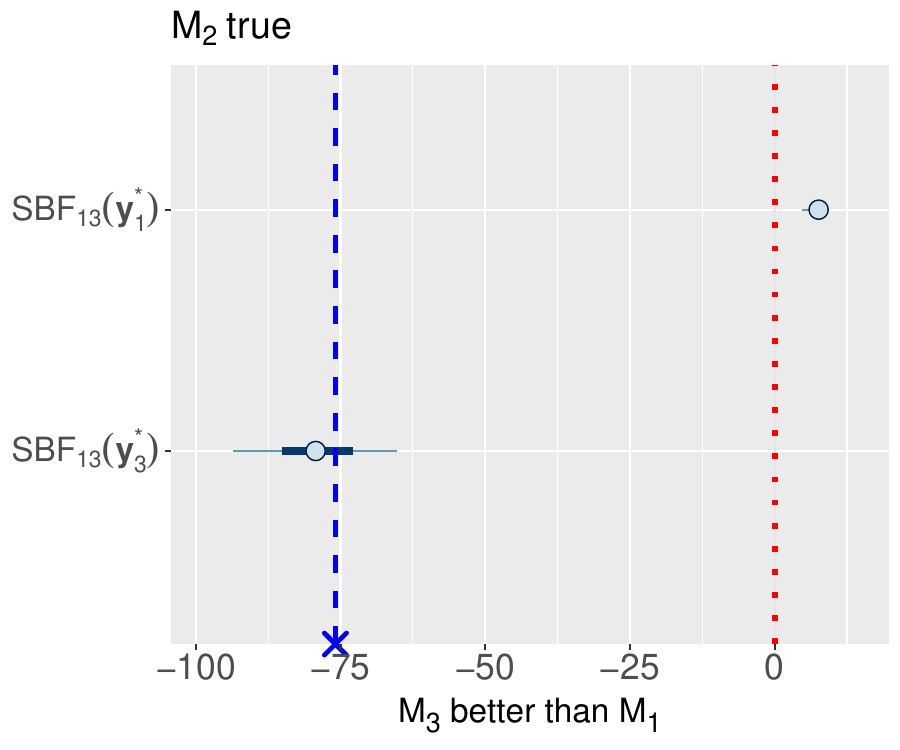}~
	\includegraphics[width=2in,height=\textheight]{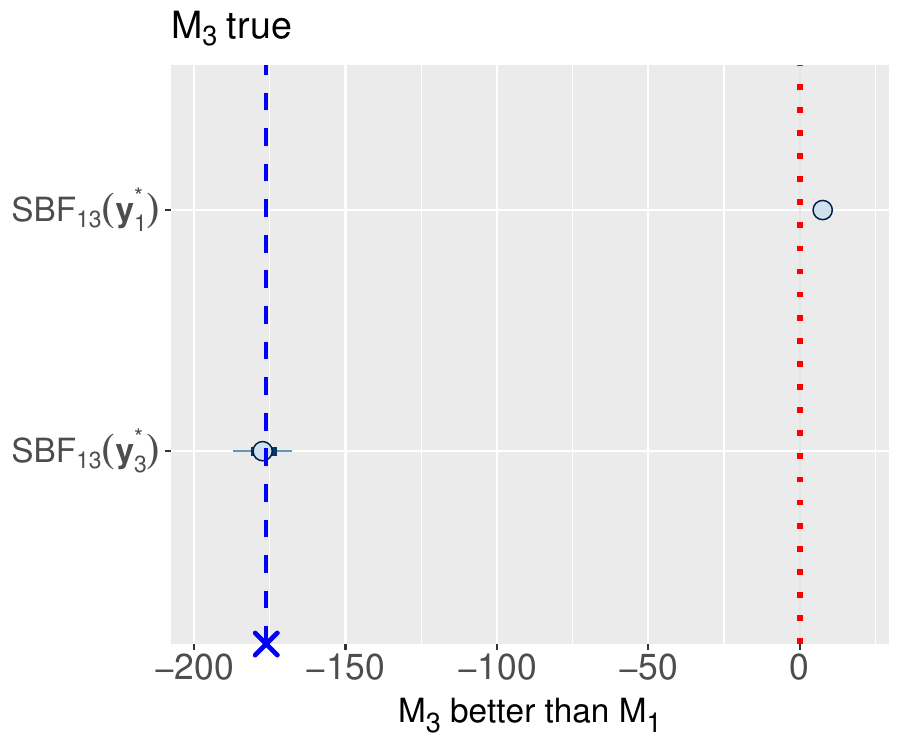}\\
	\hspace{1em} ~ 
	\parbox[h]{14.2cm}{
		\tiny \it Each column refers to the same true model; Each panel reports a log-scale pairwise comparison in terms of the 50\% (thicker lines) and 90\% (thinner lines) credible intervals for  the SBFs under the posterior-predictive replications,  $\text{SBF}_{\ell k}(\bm{y}^*)$, and the observed log-BFs evaluated on the original data (vertical dashed  plus a cross). The dotted line correspondents to zero log-BF value.}
	\caption{Error bars of the posterior predictive distributions of SBFs (in log-scale) for the nested linear regression model comparison with $g$-priors under three simulated scenarios of Section \ref{sec:gprior-nested}. 
		\label{fig:bf_linear_g}}
\end{figure}

Figure \ref{fig:bf_linear_g} depicts three simulated scenarios, where the columns denote the true model, one among $\mathcal{M}_1$, $\mathcal{M}_2$, or $\mathcal{M}_3$, and each row returns one of the three possible pairwise comparisons. Specifically,

\begin{align*}
	\begin{cases}  
		\bm{y} =  \  \beta_0 + \beta_1 \bm{x} + \bm{\epsilon} & \mathcal{M}_1\\
		\bm{y} =  \  \beta_0 + \beta_1 \bm{x} + \beta_2 \bm{x}^2 + \beta_3 \bm{x}^3 + \bm{\epsilon} & \mathcal{M}_2\\
		\bm{y} =  \  \beta_0 + \beta_1 \bm{x} + \beta_2 \bm{x}^2 + \beta_3 \bm{x}^3 + \beta_4 \bm{x}^4 + \beta_5 \bm{x}^5 +   \bm{\epsilon} & \mathcal{M}_3,
	\end{cases}
\end{align*}
where, for each scenario, we generate $n=100$ data points under the ``true'' model,  with $\bm{x}$ sampled from a standard Gaussian distribution, true parameter values $\bm{\beta} =(3, 0.3, 0.5, 0.7,1.2, -0.4)$, and $\epsilon_i \sim \mathcal{N}(0, 0.7^2)$, for $i=1,2,\ldots,n$. Each column reports the true model,  each panel reports a log-scale pairwise comparison in terms of the 50\% (thicker blue lines) and 90\% (thinner blue lines) credible intervals for the log-SBFs under the posterior-predictive replications,  $\text{SBF}_{k \ell}(\bm{y}^*)$ for models ${\cal M}_k$ and  ${\cal M}_\ell$, and the observed log-BFs evaluated on the original data, denoted by the vertical dashed blue line and a crossed mark; the dotted red line is in correspondence of zero.   We indicate with $C_{ij}$, for $i,j \in \{1,2,3\}^2$, the nine panels---from left to right, and from top to down. Table I in the supplementary material (e-Appendix B.1) resumes the results in Figure \ref{fig:bf_linear_g} and the final choice by using our Algorithm \ref{algo_bf}.

Overall, reliable support for a model requires that the posterior predictive distribution of the SBF under that model satisfies both compatibility and dominance, while at least one of these properties fails at the posterior predictive distribution of SBF under the competing model. The simulation results can be summarized as follows: 
(i) when the true model is nested within an overparametrized alternative, it satisfies both properties, whereas the larger model fails in dominance; 
(ii) when the true model formulation is nested in both candidate models, the simplest one satisfies both properties while the overparametrized fails in dominance (\lq\lq Occam's rasor\rq\rq{}); 
(iii) when a model is nested and compared to the true model, then the predictive distribution of the SBF under that model satisfies both properties, while the competing nested model will fail in terms of compatibility; 
(iv) when two nested models are compared and the true model is nested within the overparametrized model but not in the smaller one, then the predictive distribution of the SBF under the overparametrized satisfies both properties, while the competing nested model will fail in terms of compatibility; 
(v) when  two nested models are compared which are nested in the true model, then the predictive distribution of the SBF under the overparametrized satisfies both properties, while the competing nested model will fail in terms of compatibility. A detailed discussion of each simulation scenario is provided at e-Appendix B.1 in the supplementary material.

\subsection{Non-nested linear regressions}
\label{sec:gprior_nn}

Assume now to deal with three non-nested regression models as follows:

\begin{align*}
	\begin{cases}  
		\bm{y} =  \  \beta_0 + \beta_1 \bm{x} + \bm{\epsilon} & \mathcal{M}_1\\
		\bm{y} =  \  \beta_0 + \beta_1 \bm{z} +  \bm{\epsilon} & \mathcal{M}_2\\
		\bm{y} =  \  \beta_0 + \beta_1 \bm{u} + \beta_2 \bm{u}^2 +   \bm{\epsilon} & \mathcal{M}_3,
	\end{cases}
\end{align*}
with $\bm{x}, \bm{z}, \bm{u}$ sampled from a standard Gaussian, a Poisson with mean equal to two, and a Gamma  distribution with $a=b=2$ and mean equal to one, respectively, where $\bm{\beta}_1=(3,0.3)$ in $\mathcal{M}_1$, $\bm{\beta}_2 = (3, 0.9)$ in $\mathcal{M}_2$, and $\bm{\beta}_3= (3, -0.5, 0.7)$ in $\mathcal{M}_3$ and $\epsilon_i \sim \mathcal{N}(0, 0.7^2)$ for $i=1,2,\ldots, n$. 

The results are depicted in Figure~\ref{fig:bf_linear_g_nn} 
and reported in Table~I of the e-Appendix B.2 in the supplementary material. 
The overall pattern mirrors that of Figure~\ref{fig:bf_linear_g}: 
when the true model is included in the comparison, it is 
consistently identified as the best model, with the competing 
model failing either in dominance or compatibility. 
A notable exception arises in Panel~$C_{22}$, where the SBF 
of the true model ${\cal M}_2$ fails in compatibility, suggesting 
potential model misspecification despite a large observed 
$\log \text{BF} = 57$. When neither candidate corresponds to 
the true data-generating mechanism, both models fail in 
dominance, indicating that the two models are not separable. 
Predictive $p$-values and stochastic dominance probabilities 
are reported in Table~II of the Supplementary Material.

Finally, as an additional sensitivity check, Table III in e-Appendix B.3 of the supplementary material reports the percentage of which every property is satisfied for each of the nine panels for the nested and non-nested comparisons of Sections \ref{sec:gprior-nested} and \ref{sec:gprior_nn}. 
These percentages are obtained by replicating the simulations over 100 datasets for each of the 18 different scenarios under consideration. Overall, there is strong agreement between the patterns shown in Figures \ref{fig:bf_linear_g} and \ref{fig:bf_linear_g_nn} and their behavior across replicated datasets, indicating stability of the results.

\begin{figure}
	\centering
	\includegraphics[width=2in,height=\textheight]{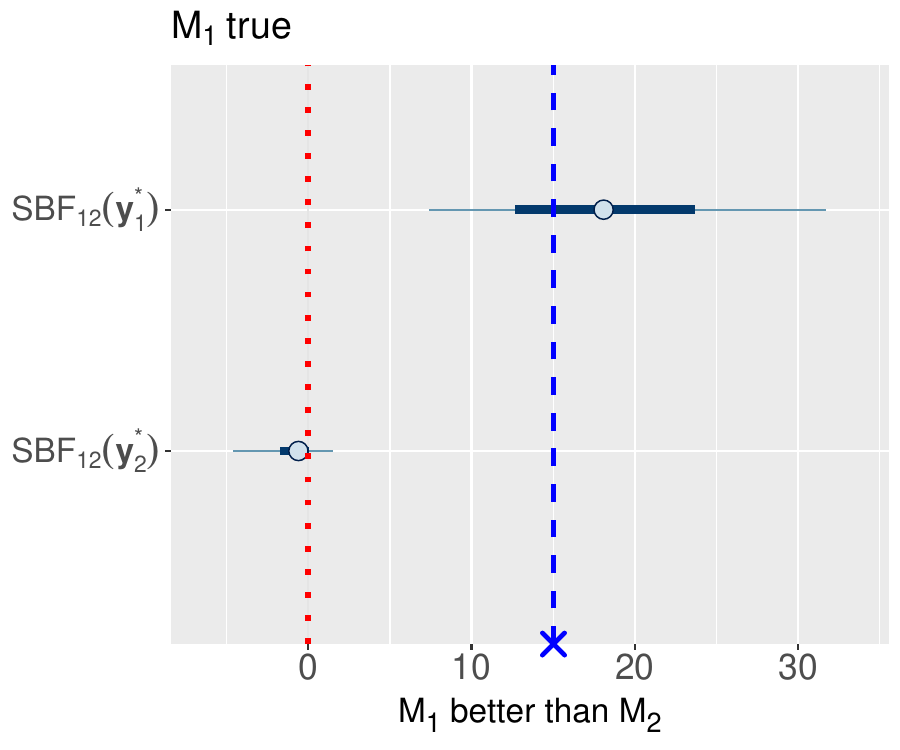}~
	\includegraphics[width=2in,height=\textheight]{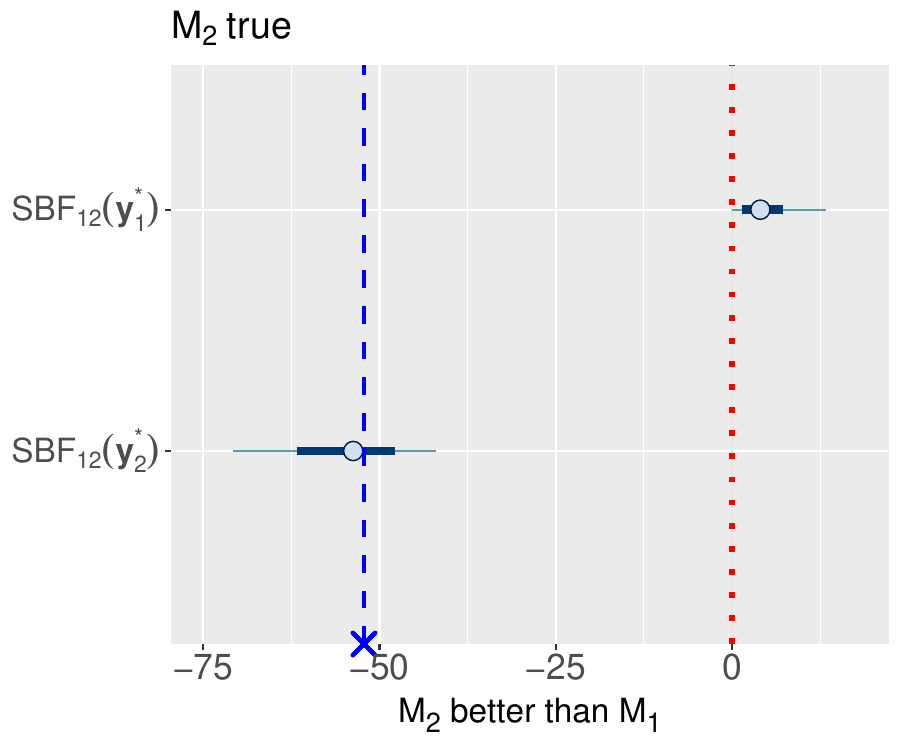}~
	\includegraphics[width=2in,height=\textheight]{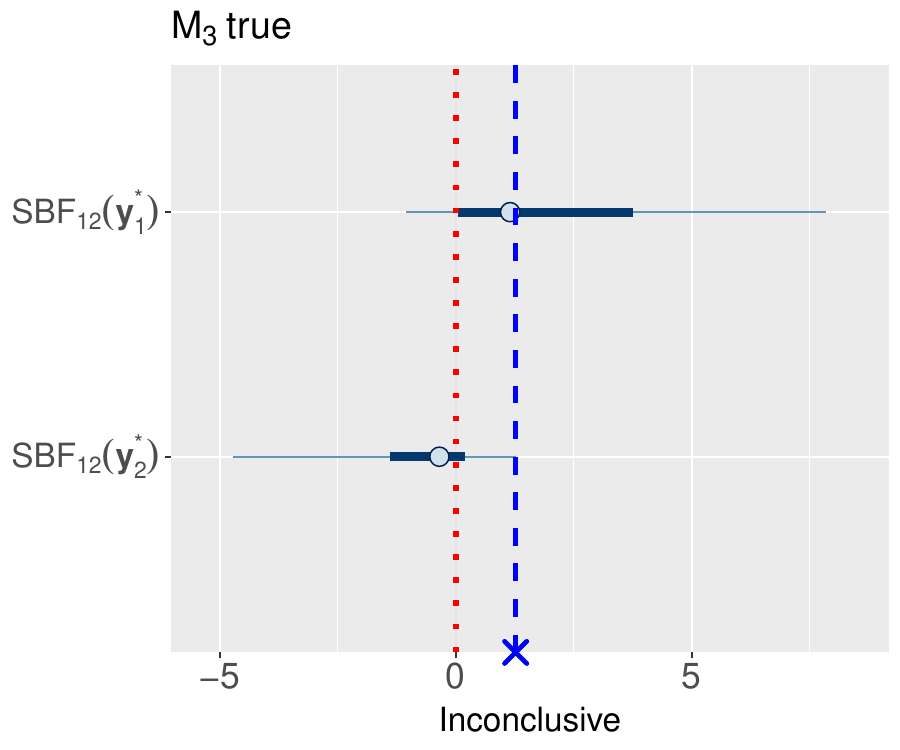}\\
	\includegraphics[width=2in,height=\textheight]{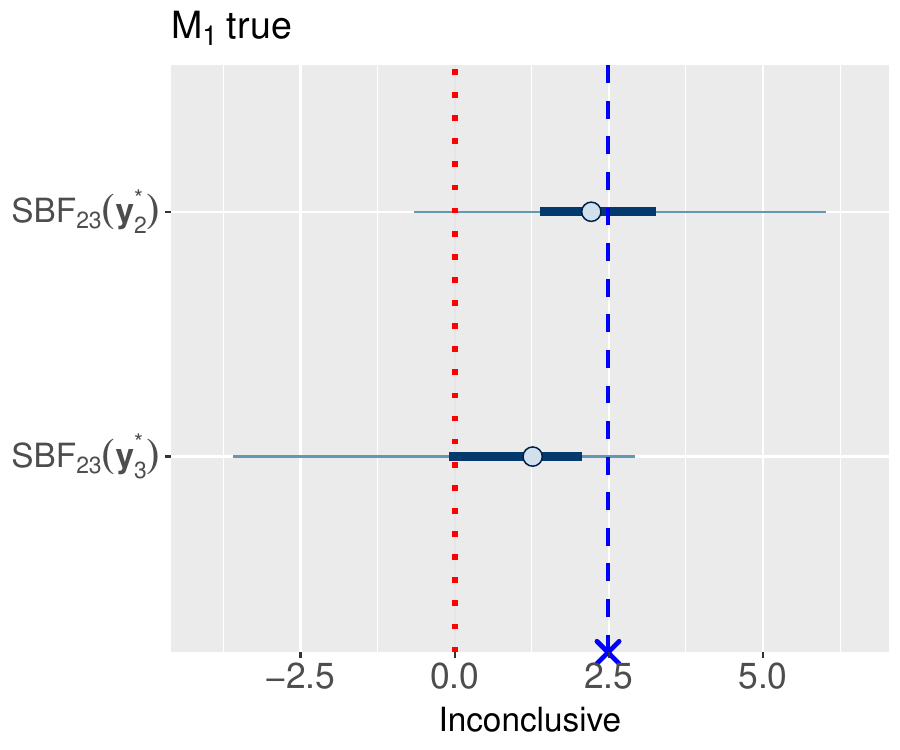}~
	\includegraphics[width=2in,height=\textheight]{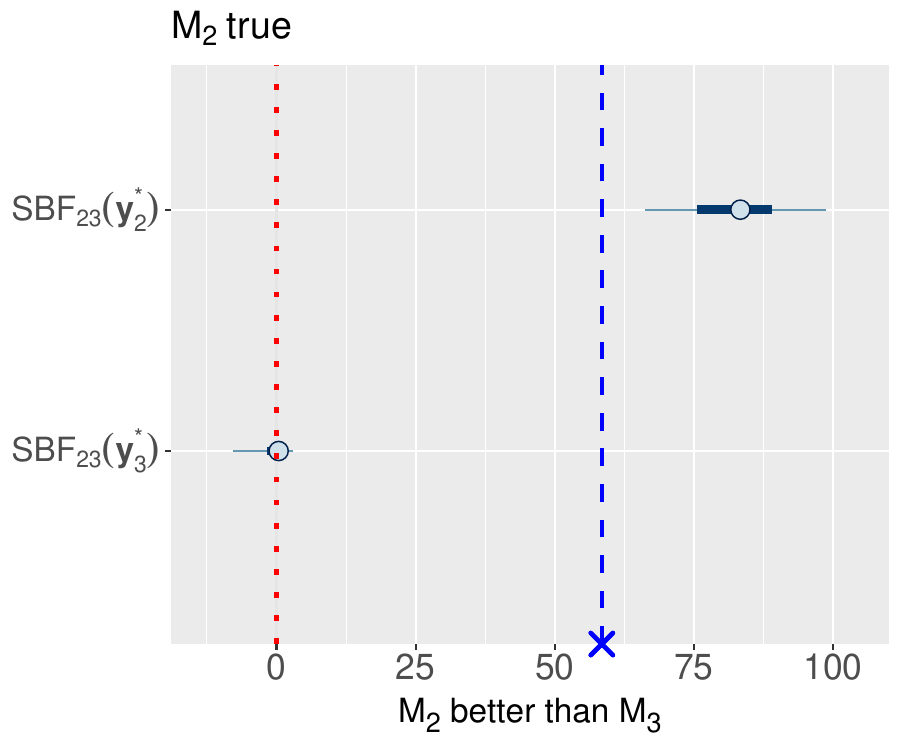}~
	\includegraphics[width=2in,height=\textheight]{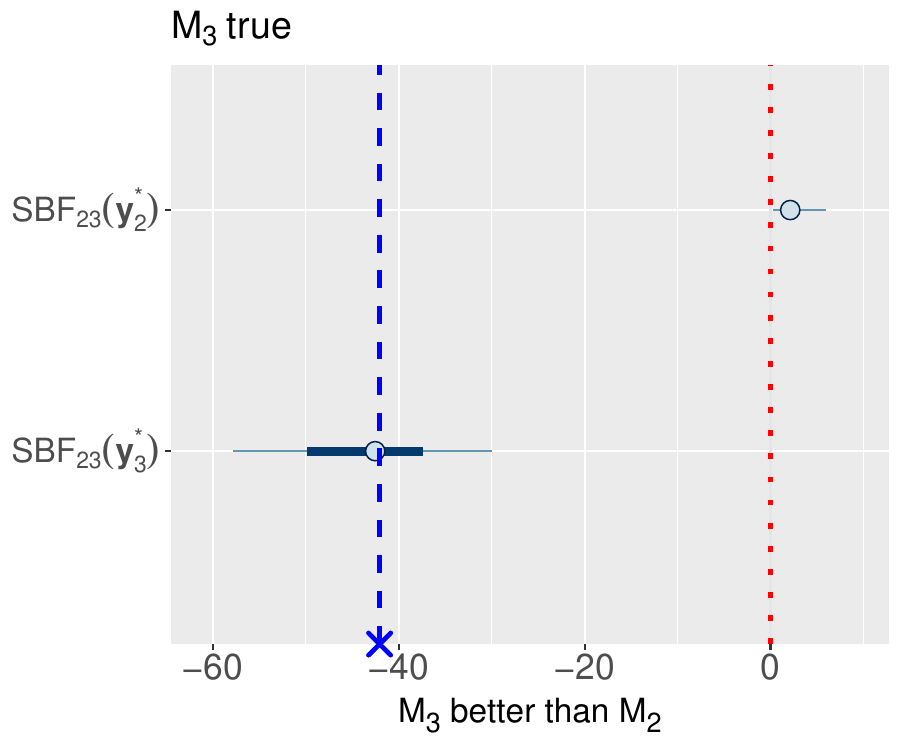}\\
	\includegraphics[width=2in,height=\textheight]{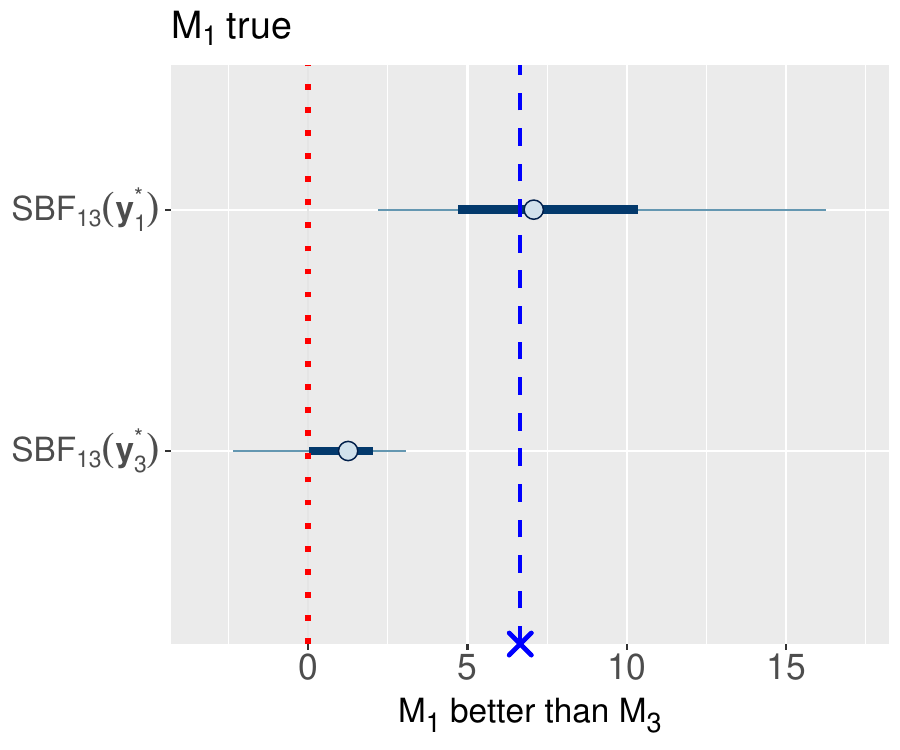}~
	\includegraphics[width=2in,height=\textheight]{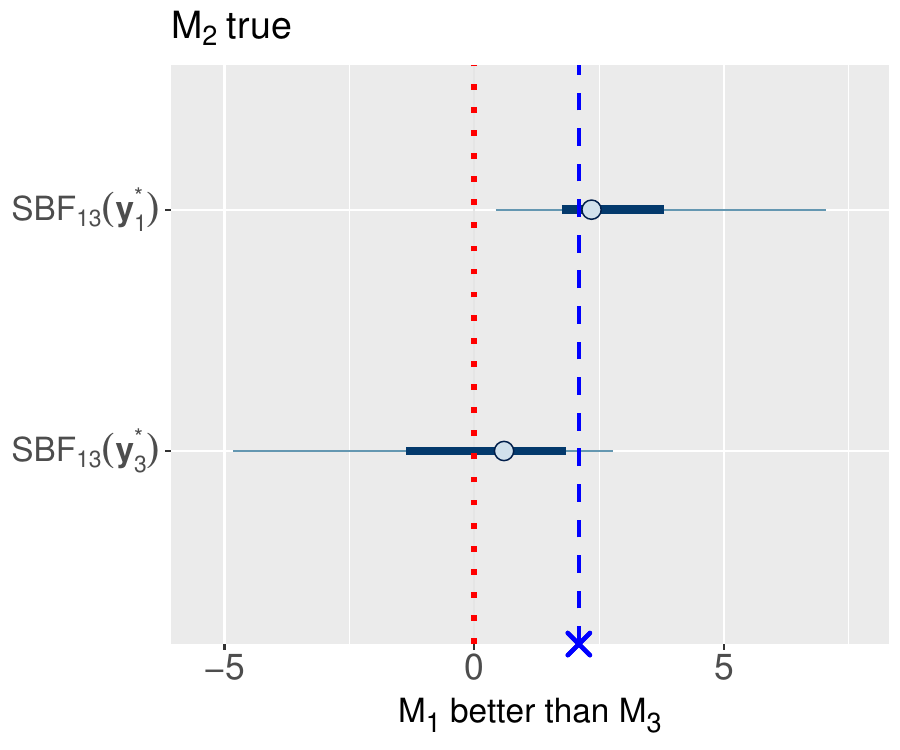}~
	\includegraphics[width=2in,height=\textheight]{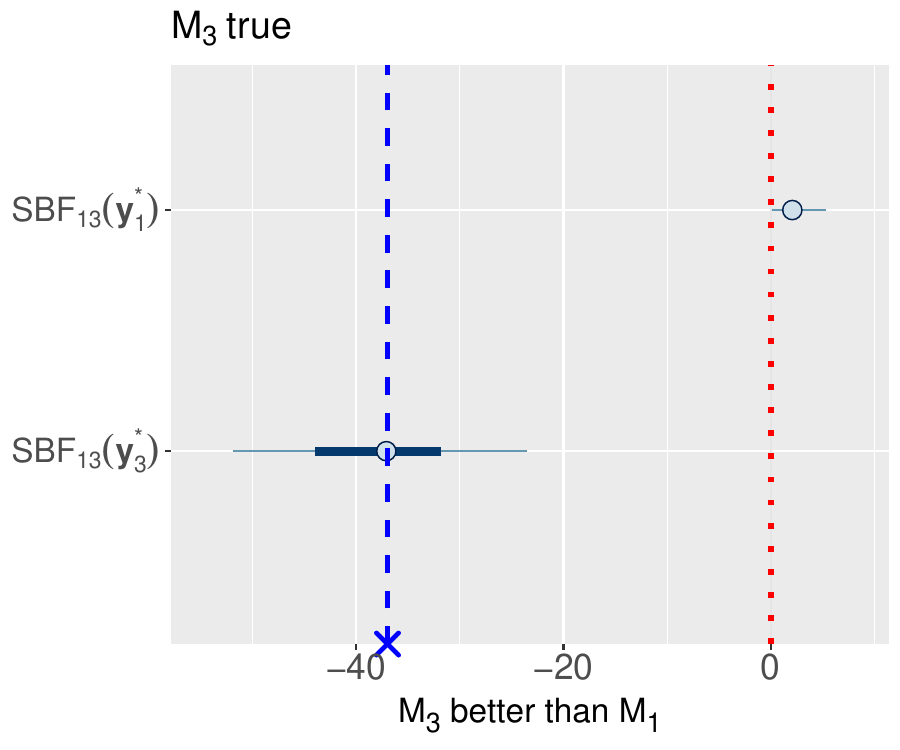}\\
	\parbox[h]{13.2cm}{
		\tiny \it Each column refers to the same true model; Each panel reports a log-scale pairwise comparison in terms of the 50\% (thicker lines) and 90\% (thinner  lines) credible intervals for  the SBFs under the posterior-predictive replications,  $\text{SBF}_{ \ell k}(\bm{y}^*)$, and the observed log-BFs evaluated on the original data (vertical dashed  plus a  cross). The dotted line correspondents to zero log-BF value.	}
	\caption{Error bars of the posterior predictive distributions of SBFs (in log scale) for the non-nested linear regression model comparison with $g$-priors under three simulated scenario  of Section \ref{sec:gprior_nn}.
		\label{fig:bf_linear_g_nn}}
\end{figure}

\subsection{Overdispersed Poisson regression: high-school attendance}
\label{sec:over}

Consider the dataset $\mathsf{Attendance}$ studied by \cite{barreto2016general} and included in the $\mathsf{R}$ package $\mathsf{mixpoissonreg}$: the aim is to assess the attendance of 314 high school junior students from two urban high schools with respect to their gender, math score and which program they are enrolled, the response variable is the number of days a student results to be absent. 
As studied by \cite{barreto2016general}, the data exhibit clear overdispersion. 
Therefore, fitting a Poisson regression would substantially underestimate the variability.
Let us now fit a Bayesian Poisson ($\mathcal{M}_1$) and a negative-binomial ($\mathcal{M}_2$) regression model by using the above response variable depending on the gender ($0=$ female, $1=$ male), the standardized math score for each student, and the three-level factor indicating the type of instructional program in which the student is enrolled. 
We use the function $\mathsf{stan\_glm}$ in $\mathsf{rstanarm}$ package, with four chains each consisting of 2000 iterations each. 
Algorithmic convergence is monitored through the Gelman-Rubin statistic $\widehat{R}$.

\begin{figure}[h!]
	\centering
	\includegraphics[width=3in,height=\textheight]{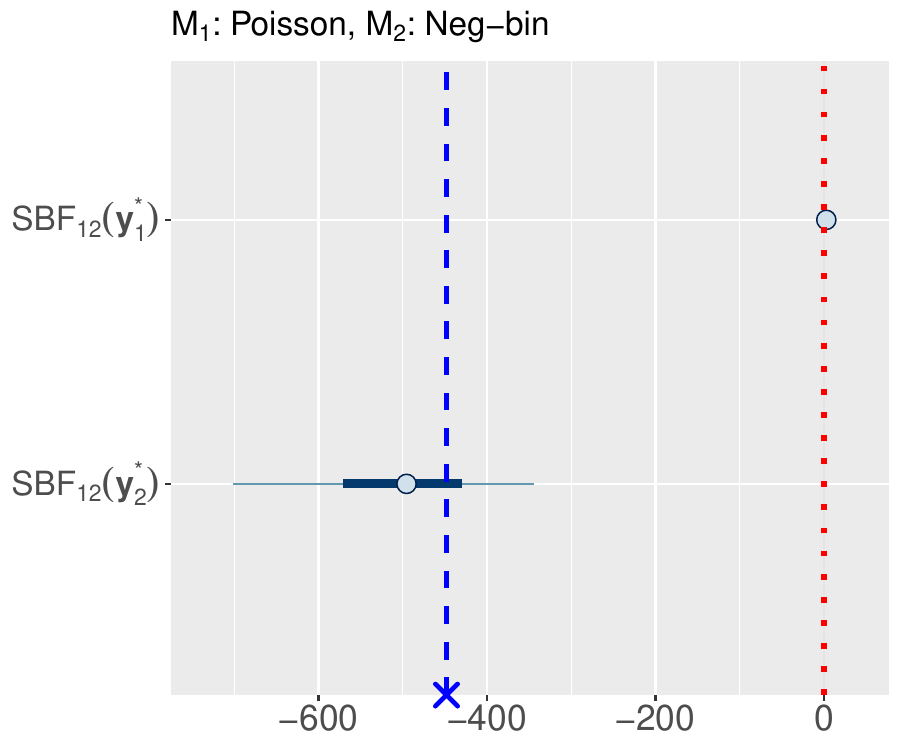}~
	\includegraphics[width=3in,height=\textheight]{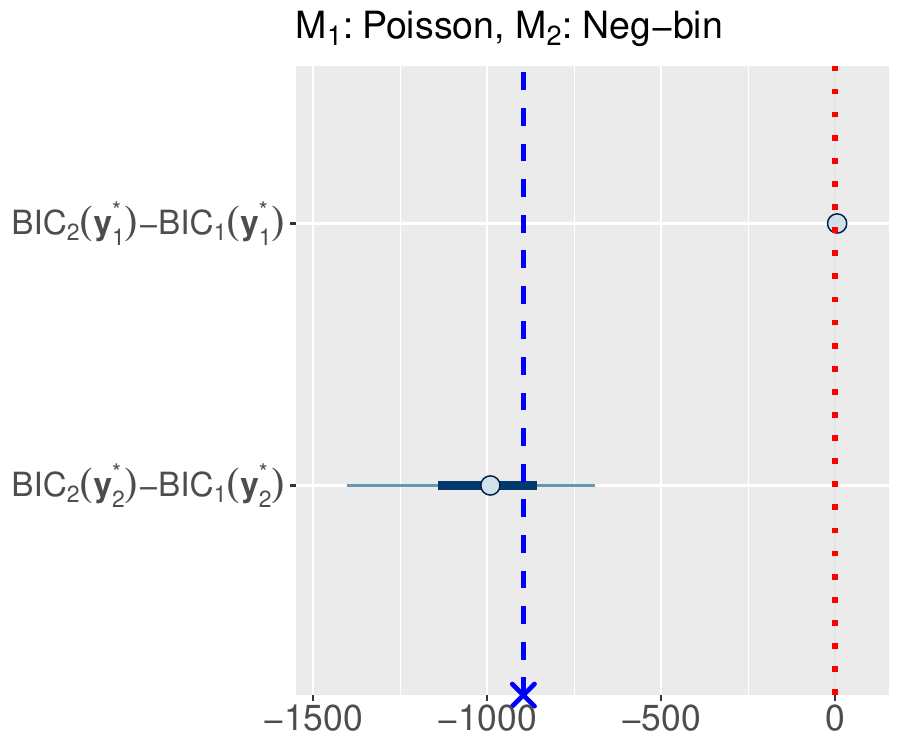}
	\parbox[h]{13.2cm}{
			\tiny \it 
			\textbf{\textit{Left Panel}}: {\textit{ Thick lines}}: 50\% posterior credible intervals; 
			{\textit{ Thin lines}}: 90\% of the  log-scale SBFs under the posterior-predictive replications of each model; 
			{\textit{ Dashed  line}}: observed BF; 
			{\textit{ Dotted linee}}: zero log-BF -- the two models are equivalent.   
			\textbf{\textit{Right Panel}}: Corresponding values for BIC differences between the two models.}
	\caption{Error bars of the posterior predictive distributions of SBFs (in log scale) for {\tt Attendance} dataset testing for overdispersion presented in Section \ref{sec:over}. 
			\label{fig:bf_pois} }
\end{figure}

Figure \ref{fig:bf_pois} reports in the left panel the 50\% (thick blue lines) and 90\% (thin blue lines) credible intervals  for the  log-scale SBFs under the posterior-predictive replications,  $\text{SBF}_{12}(\bm{y}^*_{\mathcal{M}_1})$, $\text{SBF}_{12}(\bm{y}^*_{\mathcal{M}_2})$ and the observed BF evaluated on the original data (vertical dashed blue line). 
The dashed red line highlights the zero value, corresponding to the case where the two models are considered equivalent.
It is evident that the negative-binomial model, $\mathcal{M}_2$ outperforms the Poisson model. 
Specifically, the $SBF_{12}$ under   the negative binomial model  (${\mathcal M}_2$)  indicates clear dominance of this model and it is compliant with the observed Bayes factor, while, under the Poisson model, $SBF_{12}$ fails to meet both of these criteria. 
Similar is the picture if we consider the BIC differences which may serve as a rough proxy for the Bayes factor.

\section{Discussion}
\label{sec:disc}

In this paper, we have introduced the notion of stochastic Bayes factors (SBF) as a push-forward measure obtained by transferring a measure such as the BF from the observed to the generated predictive data. Although the use of BFs and the consequent model discrimination on predictive replicates is not new in the literature \citep{garcia2005calibrating}, a general theoretical formulation is of crucial interest given the intense and informal use of the SBFs in settings such as simulation-based calibration \citep{schad2022workflow} and prior's elicitation \citep{demartino2025eliciting}. For this reason, we provide the notions of compatibility and dominance which accompany the definition of SBFs; moreover, relying on these foundational properties, we propose a novel algorithmic procedure for model discrimination using predictive samples from both the competing models.  The application on simulated and real data confirms the effectiveness of the proposed methodology, especially if compared with BFs evaluated on prior-predictive samples. In general, this procedure is likely to be beneficial in the wide setting of model selection and refreshes the notion of BF by acknowledging an intrinsic posterior uncertainty in a solid theoretical fashion.

Many points of criticism remain open. First of all, this procedure actually uses the data twice, one for computing the posterior and the other for sampling hypothetical replicates from the posterior-predictive distribution of each model. To dilute this double use of the data,  future research should focus on a predictive version of the SBFs, by possibly using a portion of the data as training set and the other portion as validation/test set, similarly to how intrinsic Bayes factors \citep{berger1996intrinsic} work. 

Second, there could be some gray situations of indecision or weak evidence where model discrimination is unclear; 
for instance when the observed BF provides strong evidence in favor of a model, but not to the extent which is expected by the SBF under that model's predictive distribution. 

Third, it would be very interesting to explore connections between the posed SBFs and cross-validation techniques, such as the leave-one-out CV: a deeper investigation is likely to connect pure Bayesian with deep/machine learning modern tools. 

Finally, in an earlier version of this manuscript, we have additionally examined the notion of separability between the predictive distributions of the SBF under the two models under comparison. 
However, in exploring this idea, we encountered difficulties in characterizing and interpreting the behavior of SBFs in terms of separability. 
At this stage, the examination of the property of separability does not appear to provide substantial additional insight beyond the notions of dominance and compatibility on which we have finally focused.

\appendix 

\section*{Disclosure Statement}
The authors report there are no competing interests to declare.

\section*{Funding}
No funding was obtained for the reported work.

\section*{Declaration of generative AI use}

During the preparation of this work, the author(s) used Claude (Anthropic) in order to: (i) assist with aesthetic refinement of plot styling and visualization choices; (ii) independently cross-check the derivation steps in the proof of Theorem 2; and (iii) perform light proofreading of the manuscript text for language and clarity. After using this tool, the author(s) reviewed and edited all content as needed and take full responsibility for the content of the published article. All research ideas, methodology, theoretical results, and conclusions remain entirely the authors' own.

\section*{Data availability statement}

Data available within the article or its supplementary materials.

\phantomsection\label{supplementary-material}
\bigskip

\begin{center}

{\large\bf SUPPLEMENTARY MATERIAL}

\end{center}

\begin{description}
\item[Title: suppl\_mat.pdf]
It contains the technical proofs for Theorems 1 and 2 and further tables and comments for the applications in Section \ref{sec:ex}.

\item[\texttt{Attendance} data set:]
Data set contained in the \texttt{mixpoissonreg} \texttt{R} package and used in the illustration of the overdispersed Poisson regression in Section~\ref{sec:over}. 
\end{description}


\end{document}